\documentclass[a4paper,11pt]{article}
\pdfoutput=1 % if your are submitting a pdflatex (i.e. if you have
\usepackage{jcappub} % for details on the use of the package, please
\usepackage{aas_macros} % to make journal abbreviations in AAS bib entries work

\usepackage[T1]{fontenc} % if needed
\usepackage[utf8]{inputenc}

\usepackage[english]{babel}
\usepackage{amsmath}
\usepackage{amssymb}
\usepackage{MnSymbol}
\usepackage{nicefrac}
\usepackage{amsfonts}
\usepackage{dsfont}

\usepackage{natbib}
\usepackage[markup=nocolor]{changes}

\usepackage{graphicx}% Include figure files

\usepackage[normalem]{ulem}

\usepackage{dcolumn}% Align table columns on decimal point
\usepackage{bm}% bold math

\usepackage{color}
\usepackage[usenames,dvipsnames,svgnames,table]{xcolor}
\usepackage{slashed}
\usepackage{empheq}
\newcommand{\bea}{\begin{eqnarray}}
\newcommand{\eea}{\end{eqnarray}}
\newcommand{\be}{\begin{equation}}
\newcommand{\ee}{\end{equation}}

\title{\boldmath Asymptotic safety casts its shadow}

\author[a,1]{Aaron Held,\note{Corresponding author.}}
\author[b]{Roman Gold,}
\author[c,a]{Astrid Eichhorn}

\affiliation[a]{Institut f\"ur Theoretische Physik, Universit\"at Heidelberg,
\\
Philosophenweg 16, 69120 Heidelberg, Germany}
\affiliation[b]{Institut f{\"u}r Theoretische Physik,
Johann Wolfgang Goethe-Universit\"at,
\\
Max-von-Laue-Stra{\ss}e 1, 60438 Frankfurt, Germany}
\affiliation[c]{CP3-Origins, University of Southern Denmark,
\\
Campusvej 55, DK-5230 Odense M, Denmark}

\emailAdd{held@thphys.uni-heidelberg.de}
\emailAdd{gold@itp.uni-frankfurt.de}
\emailAdd{eichhorn@sdu.dk}

\abstract{We set out to bridge the gap between 
regular black-hole spacetimes and observations of a black-hole shadow by the Event Horizon Telescope. We explore modifications of
spinning and non-spinning black-hole spacetimes inspired by asymptotically safe quantum gravity which features a scale dependence
of the Newton coupling. As a consequence, the predictions of our model, such as the shadow shape and size, depend on one free
parameter determining the curvature scale at which deviations from General
Relativity set in. In more general new-physics settings, it can also depart
substantially from the Planck scale. In this case, the free parameter is constrained by observations, since the corresponding curvature scale is significantly below the Planck-scale.
The leading new-physics effect can be recast as a scale-dependent black-hole mass, resulting in distinct observational signatures of our model.  As a concrete example, we show that two mass-measurements, extracted from the size of the shadow and from Keplerian orbital motion of stars, allow to distinguish the classical from the modified, regular black-hole spacetime, yielding a bound on the free parameter.  
For spinning black holes, we further find that the singularity-resolving new physics puts a characteristic dent in the shadow. 
Finally, we argue, based on the underlying physical mechanism, that the effects we derive could be generic consequences of a large class of quantum-gravity theories.}

\begin{document}
\maketitle
\flushbottom

%=======================================================================================================

\section{Motivation}
The true nature of black holes, either those located at essentially all 
centres of galaxies \cite{Cattaneo2009} or their stellar mass
counterparts \cite{Remillard2006} is an exciting question that brings us
to the limit of (if not beyond) our understanding of the inner workings
of gravity.
Quantum gravity is
expected to strongly modify the geometry of spacetime in the interior of
black holes, where classical solutions feature singularities.
Quantum-gravity effects in black-hole spacetimes have been explored,
e.g.~in 
\cite{tHooft:1984kcu,Ashtekar:1997yu,Gauntlett:1998fz,Horowitz:1998pq,1999alfiomartin,br00,BjerrumBohr:2002ks,Fidkowski:2003nf,Nicolini:2005gy,Nicolini:2008aj,He:2008ku,Gambini:2008dy,Gambini:2013ooa,Rovelli:2014cta}. Moreover,
proposals for singularity-free black-hole-like spacetimes have been made, e.g.~in
\cite{Bardeen,Dymnikova:1992ux,Dymnikova:1996ux,Hayward:2005gi}.  In
these cases, the metric is strongly modified close to the classical
singularity, but effects can persist even outside the horizon.

 The possibility to directly image a black hole
 with the technique
of Radio Very-Long-Baseline-Interferometry
\cite{Falcke:1999pj,Doeleman:2008qh,paper1,paper2,paper3,paper4,paper5,paper6} is an exciting and
promising development in gravitational physics. 
Zooming in on the
horizon, will there be detectable imprints of the fundamental quantum
structure of spacetime? According to a 
standard power-counting
argument such
signatures are highly suppressed by a power of $M_{\rm Pl}/M$ where
$M$ is the mass of the black hole and $M_{\rm Pl}$ the Planck mass, at
least for astrophysically plausible black holes with masses between
several tens to several billion solar masses. So, the conservative answer
would be ``no''. Yet, the possibility of quantum-gravity
effects at the horizon has been raised, e.g.~in \cite{Almheiri:2012rt,Haggard:2016ibp,Giddings:2017jts,Compere:2019ssx,Giddings:2019ujs}.
In particular, for extended objects like black-hole horizons, a simple power-counting argument based on the \emph{local} curvature might be misleading.
Moreover, given that singularities plague black-hole spacetimes in General Relativity, some form of new physics must exist to resolve these, not necessarily being quantum gravity.  This motivates us to explore quantum-gravity inspired models while keeping the scale at which deviations from (General Relativity) GR set in as a free parameter.
However, in order to confront theoretical models with the data, we
first need to know what we are looking for. In the classical case, the image of a black hole has first been studied, e.g.~in \cite{Hilbert1917,vonLaue1921,Bardeen1973,Luminet1979}.
To construct the quantum analogue, potential
consequences of quantum gravity for the spacetime geometry close to the
horizon have yet to be understood even qualitatively.  

Game-changing detections of gravitational waves \cite{Abbott:2016blz} and the groundbreaking observations of the first image of  M87* by the EHT \cite{paper1,paper2,paper3,paper4,paper5,paper6} now enable observational probes of the metrics of the most compact objects in our universe. This provides access to a completely different regime of gravity than the well-tested weak-field regime, e.g.~in the solar system, where GR has been tested extensively \cite{Will:2014kxa}. In contrast, both the origin of the detected gravitational-wave signals as well as the radio-emission detected by the EHT lie in the strong-field regime, where curvature effects become significant. A clear prerequisite for tests of GR based on EHT observations, as advocated in \cite{Falcke:2013ola,2014ApJ...784....7B,Johannsen:2015mdd,2015ApJ...814..115P,Johannsen:2015hib,Younsi:2016azx,Johannsen:2016uoh,Psaltis:2018xkc,2018GReGr..50...42C}, is the development of model-predictions from modifications of GR, some of which might be of quantum origin.
Specifically, the observable we focus on is the shape and size of black-hole shadows, which has become visible in images of horizon-scale structures \cite{paper1, paper2, paper3, paper4, paper5, paper6}. 
A black hole by itself does not emit any light -- by construction. 
It also blocks a significant portion of light emitted behind the black hole. The resulting darkness contrasts against any otherwise bright background emission. Moreover, the strong bending of light rays by the black hole generates a luminous ring of light demarcating the shadow. The shape and size of the shadow can be calculated given a form of the metric describing the black-hole spacetime. Due to the no-hair theorem, the mass and spin of the black hole are the only two parameters determining shape and size of the shadow in GR. In this case, the black-hole shadow is nearly circular in shape even for rapidly spinning black holes \cite{Psaltis:2018xkc}. 
In alternative theories of gravity, richer structures are possible \cite{Johannsen:2010ru,Cardoso:2016ryw,CarballoRubio:2018jzw}. 
Shadows in modified theories of gravity have been studied for Chern-Simons gravity \cite{2010PhRvD..81l4045A,Okounkova:2018abo}, in Randall-Sundrum-type braneworld models \cite{Amarilla:2011fx} (also with cosmological constant \cite{Eiroa:2017uuq}), in Scalar-Tensor-Vector-Gravity \cite{Moffat:2015kva,Wang:2018prk} and tensor-vector gravity 
\cite{Vetsov:2018mld}, 
Einstein-dilaton-Gauss-Bonnet gravity
\cite{2017PhLB..768..373C,Ayzenberg:2018jip}, 
Einstein-Maxwell-dilaton-axion gravity 
\cite{2013JCAP...11..063W}, 
Einstein-Born-Infeld gravity
\cite{Atamurotov:2015xfa} and conformal gravity \cite{Mureika:2016efo}.  
Further, the shadows of super-extremal Kerr black holes \cite{Bambi:2008jg} and Kerr-Newman-NUT spacetimes \cite{2014PhRvD..89l4004G} as well as Kerr black holes with scalar hair 
\cite{2015PhRvL.115u1102C,2016PhRvD..94h4045V}
have been explored. Regular spacetimes have been studied in \cite{Abdujabbarov:2016hnw,Amir:2016cen,Tsukamoto:2017fxq,Stuchlik2019}. The shadow of non-commutative-geometry inspired black-hole spacetimes has been investigated in \cite{Wei:2015dua}.
As an alternative to separate studies within distinct gravity theories, one can parameterise a more general metric than the Kerr metric, providing a way to constrain more than one theory of gravity at the same time
\cite{Rezzolla:2014mua,Johannsen:2015pca,Konoplya:2016jvv,Younsi:2016azx}. For actual comparisons with EHT observations, it is necessary to realistically model the dynamics of infalling matter, e.g.~using GRMHD simulations \cite{Mizuno2018}.

Until now, no such predictions from asymptotically safe gravity \cite{Weinberg:1980gg,Reuter:1996cp} are available and we set out to close this gap. 
 The theory is based on the well-tested quantum-field theory framework which can successfully describe the three other fundamental interactions of nature, and extends it to include gravity. Further, it respects the observational evidence that the gravitational dynamics can be formulated purely in terms of the metric field by promoting the metric to a fundamental quantum field. Motivated by this conservative nature of asymptotically safe gravity, we explore asymptotic-safety inspired black-hole spacetimes to derive shape and size of the black-hole shadow. The key idea explored in this work consists in the onset of quantum-gravity effects and resulting modifications of the spacetime at large curvature scales. As one might expect, Planck-scale modifications of GR cannot be constrained by the EHT observations.
 
 While the specific black-hole spacetimes we analyse are motivated by an approach to quantum gravity, we take a broader  view in which the modifications of GR are not necessarily tied to the Planck scale. Remaining agnostic about the type of new physics that sources the  singularity-resolving modifications, we derive bounds on the new parameter determining the size of the deviations from GR.
 
 This paper is structured as follows. We provide an introduction to asymptotically safe quantum gravity and explain how to set up asymptotic-safety inspired black-hole spacetimes in the spherically symmetric and axisymmetric case in Sec.~\ref{sec:BHinAS}. We review how to calculate the shadow of a black hole in Sec.~\ref{sec:methods} and present our results in Sec.~\ref{sec:results}, where we also interpret the physical mechanism underlying the resulting deformation of the shadow compared to the classical case. We further broaden our view beyond quantum gravity as the new physics that leads to singularity resolution and  explain how to constrain a wider parameter space in our model. In Sec.~\ref{sec:conclusions}, we summarise our key findings and argue that qualitatively similar effects could be a generic consequence of a large class of quantum-gravity theories.

\section{Black holes in asymptotically safe gravity}\label{sec:BHinAS}
Quantum fluctuations of spacetime result in a scale-dependence of the gravitational interactions. Thus, unlike in classical gravity, the Newton coupling is not a simple constant, but changes as a function of scale due to the impact of quantum fluctuations, just like the other fundamental interactions of nature ``run" as a consequence of quantum fluctuations, (as for instance in QCD). In quantum field theories, such a scale dependence can terminate in an unphysical infinity, a so-called Landau pole, which implies a breakdown of the quantum-field theoretic description at the corresponding finite scale and requires some form of new physics. 
Under one condition, such poles can be avoided, rendering the quantum field theory a viable description up to arbitrarily microscopic scales. This condition  consists in an enhanced symmetry, namely quantum scale-invariance: In such a scale-invariant  regime, the dynamics do not depend on the scale, allowing one to zoom in up to arbitrarily small scales without running into any inconsistencies. The simplest form of quantum scale invariance is asymptotic freedom, which consists in the absence of interactions in the microscopic regime. While it plays a central role for the Standard Model of particle physics, asymptotic freedom cannot be achieved in quantised GR due to its infamous perturbative nonrenormalisability \cite{tHooft:1974toh,Goroff:1985sz,vandeVen:1991gw}. Historically, this result led to the development of alternative descriptions of quantum gravity, including, e.g.~string theory and loop quantum gravity. Yet, just as for other fundamental interactions, an alternative to asymptotic freedom could exist also for a quantum field theory of the metric, namely asymptotic safety. In an asymptotically safe setting, dimensionless coupling strengths become constant beyond the transition scale to the scale-invariant regime, realising quantum scale invariance in the presence of residual interactions. 
For dimensionful interaction strengths, such as the Newton coupling, quantum scale-invariance implies that they scale canonically with the Renormalisation Group (RG) scale $k$. This ensures that the dimensionless counterpart of these interactions (constructed by multiplication with appropriate powers of $k$, is constant instead of changing with $k$) just as it has to be in a scale-invariant setting, see Eq.~\eqref{eq:GN} for the Newton coupling.

Thus, the key idea of asymptotically safe gravity \cite{Weinberg:1980gg} is that quantum scale-symmetry holds beyond a microscopic transition scale, typically assumed to be the Planck scale. This requires quantum fluctuations to be antiscreening, such that the dimensionless counterpart  of the Newton coupling becomes constant beyond the Planck scale, $g(k)= G_N(k) k^2 = \rm const$. For the dimensionful Newton coupling, the antiscreening by quantum-gravity fluctuations results in a fall-off
\be
G_N(k) = \frac{g_{\ast}}{k^2},\label{eq:GN}
\ee
beyond the transition scale to the fixed-point regime. Herein $g_{\ast}$ is the asymptotically safe fixed-point value.
A consequence of Eq.~\eqref{eq:GN} is an effective weakening of gravity in the far ultraviolet.
 Heuristically, this is just what one would expect for a quantum theory of gravity that is capable of resolving the singularities present in classical gravity.
  
 Compelling indications for the existence of an asymptotically safe regime in quantum gravity have been found \cite{Reuter:2001ag,Lauscher:2001ya,Litim:2003vp,Codello:2008vh,Benedetti:2009rx,Dona:2013qba,Falls:2013bv,Becker:2014qya,Gies:2016con,Christiansen:2017bsy,Falls:2018ylp,Eichhorn:2018ydy}, see \cite{Percacci:2017fkn,Reuter:2019byg} for introductions.  Unexpected potential  implications for particle-physics phenomenology have been  found in \cite{Shaposhnikov:2009pv,Eichhorn:2017ylw,Eichhorn:2018whv},
for recent reviews containing a discussion of open questions see \cite{Eichhorn:2017egq,Eichhorn:2018yfc}.

The full dynamics of gravity in the far UV in asymptotic safety includes terms beyond the Einstein-Hilbert action \cite{Benedetti:2009rx,Gies:2016con,Christiansen:2017bsy,Falls:2018ylp}. 
For the purposes of our paper, we will limit ourselves to a simplified 
 setting and explore the scale-dependence of the Newton
coupling for black-hole spacetimes as an expected leading-order effect of asymptotically safe gravity. 
The effect of a cosmological constant has been explored in \cite{Koch:2014cqa,ko14,Pawlowski:2018swz,Adeifeoba:2018ydh}, where an RG improvement considering a canonical scaling of the cosmological constant in the fixed-point regime leads to the reintroduction of a singularity. Yet, an IR value of the cosmological constant, as required by cosmological observations, is compatible with a vanishing fixed-point value, such that the cosmological constant might not play a role for the structure of the black-hole spacetime in the vicinity of the classical singularity, as analysed in  \cite{Adeifeoba:2018ydh}. We follow \cite{Adeifeoba:2018ydh} in constructing RG-improved Schwarzschild and Kerr spacetimes and treat the cosmological constant as an IR scale that can be neglected in our setting.
The impact of higher-order curvature terms, as expected from asymptotic safety, on the black-hole shadow will be studied elsewhere \cite{HGE}.
In exploring the impact of a scale-dependent Newton coupling on black-hole shadows, we make the key assumption that those indications for asymptotic safety in gravity collected in the Riemannian regime (with the exception of \cite{Manrique:2011jc}) are applicable to the Lorentzian regime.

\begin{figure}
\centering
\includegraphics[width=0.6\linewidth]{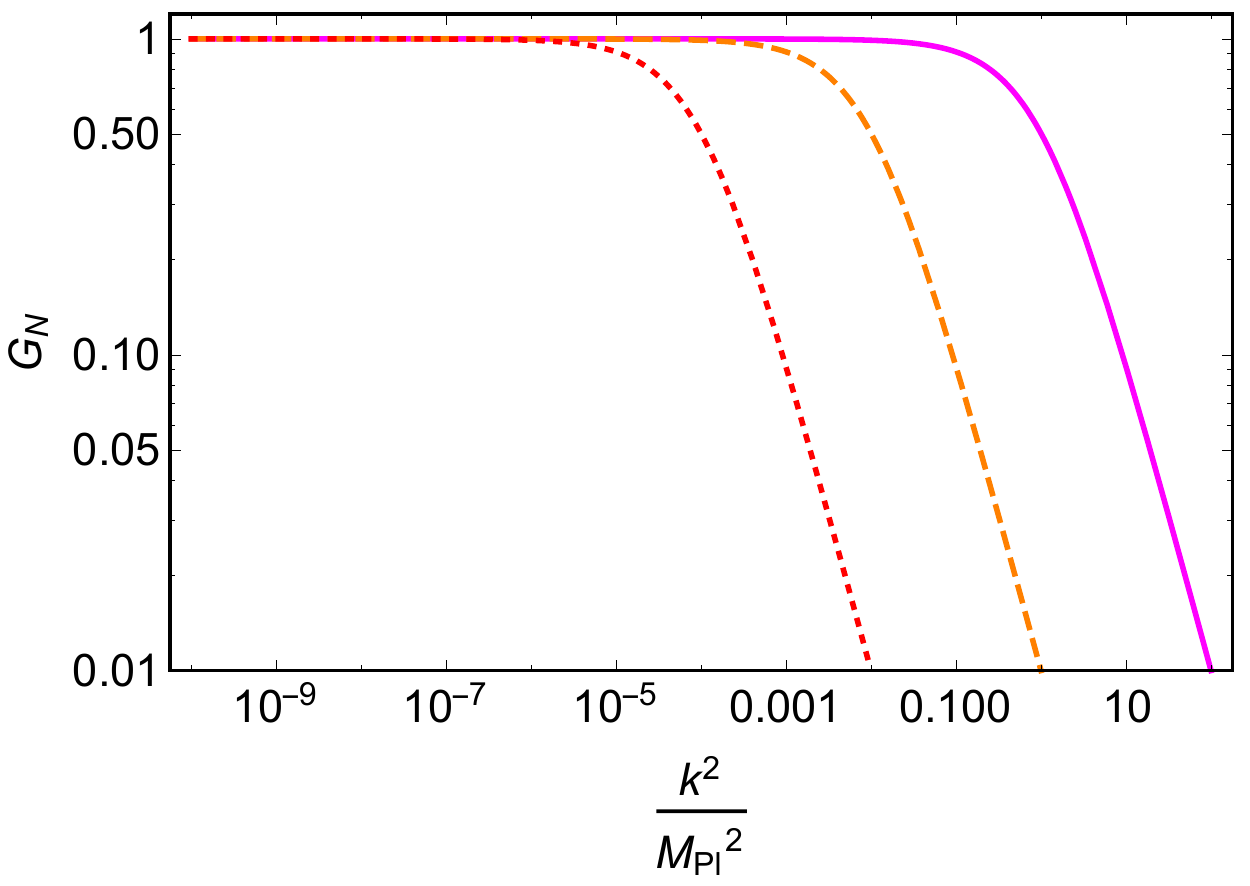}
\caption{\label{fig:GNrun} Running Newton coupling in units of the classical Planck mass for $\gamma=1$ (continuous magenta line), $\gamma=10^2$ (orange dashed line) and $\gamma=10^4$ (red dotted line). The classical regime with $G_N=\rm const$ and the quantum regime with $G_N \sim k^{-2}$ are separated by the transition scale $\gamma^{-1} M_{\rm Pl}$.}
\end{figure}

To upgrade the classical black-hole spacetimes to modified, ``asymptotic-safety inspired" black-hole spacetimes, we focus on the scale dependence of the Newton coupling, generated by quantum fluctuations
\be
G_N(k) = \frac{G_0}{1+\gamma\, G_0 k^2}.
\ee
Here, $G_0$ is the classical value of the Newton coupling, and $\gamma>0$ is the inverse dimensionless fixed-point value ($\gamma = g_{\ast}^{-1}$), i.e.~the square of the inverse transition scale to the fixed-point regime, measured in Planck units. In a simple approximation, the scale comes out as roughly the Planck scale, but upgrades including the impact of quantum-fluctuations of matter show that $\gamma^{-1}$ can vary  \cite{Dona:2013qba}.
$k$ is the Renormalisation Group scale, i.e.~an energy scale that can loosely be interpreted as the inverse resolution scale of the theory: For low $k$, one probes the classical regime of the theory, whereas one enters the quantum regime for $k^2 G_0>\gamma^{-1}$. In other words, $\gamma^{-1}$ is the transition scale to the quantum regime in units of the classical Planck mass, cf.~figure~\ref{fig:GNrun}. For the remainder of this paper, we treat $\gamma$ as a free parameter.

Hints that the fall-off of the Newton coupling due to antiscreening quantum fluctuations in the UV could indeed lead to singularity-free black-hole spacetimes have been found in \cite{1999alfiomartin,br00,2006alfiomartin,Falls:2010he,Falls:2012nd,Litim:2013gga,torres14,ko14,Koch:2014cqa,BKP,torres15,Kofinas:2015sna,Bonanno:2017zen,Pawlowski:2018swz,Adeifeoba:2018ydh,Platania:2019kyx,Bosma:2019tbl}. In these works, the key idea is to investigate an effective metric that should be understood as the expectation value of the metric in the quantum theory. Starting from the classical metric, e.g.~Schwarzschild or Kerr, the RG-improved metric is constructed following two steps. Firstly, the Newton coupling in the metric is replaced by its scale-dependent counterpart which encodes the effect of quantum fluctuations of gravity. Secondly, the Renormalisation Group scale $k$ that the Newton coupling depends on is identified with a physical scale of the classical spacetime. This results in a modified metric, which is a solution to the Einstein equations with an effective energy-momentum tensor. The interpretation of the latter is that it encodes the quantum-gravity contributions at an effective level.

We stress that this procedure it not a strict derivation from asymptotically safe gravity. Instead, it makes two crucial assumptions: (i) The full dynamics of the theory is truncated to an Einstein-Hilbert term. (ii) Quantum effects are included by taking into account the RG scale dependence of the Newton coupling, and making the assumption that a physical scale of a classical black-hole spacetime can be identified with the RG scale. 
Overall this results in asymptotic-safety inspired models for regular black-hole spacetimes.

\subsection{Improved Schwarzschild spacetime}
\begin{figure}
\centering
\includegraphics[width=0.6\linewidth]{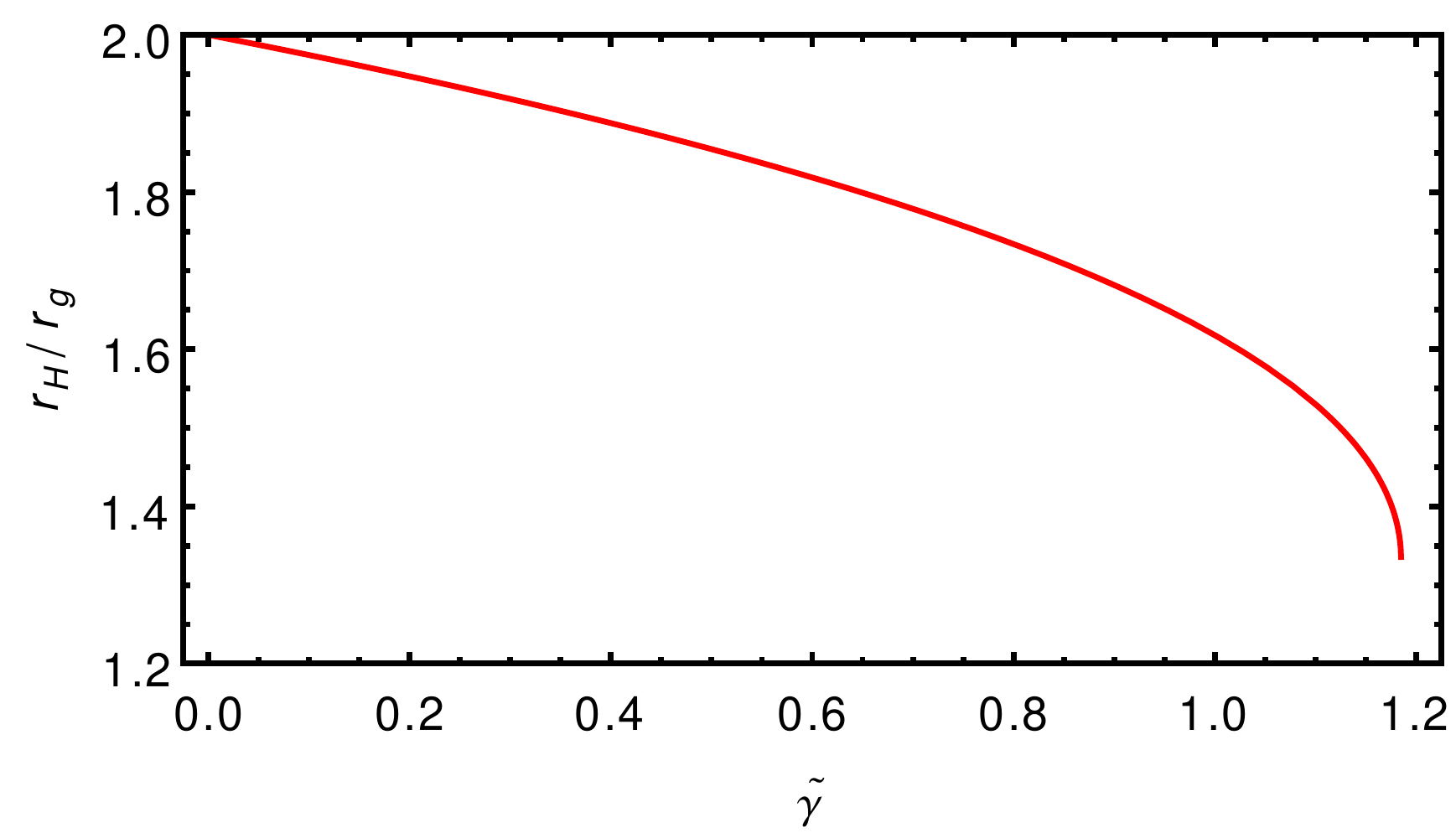}
\caption{\label{fig:rgimprovedhorizon} We show the radial coordinate of the horizon as a function of $\tilde{\gamma}$, see Eq.~\eqref{eq:rhSchw}.}
\end{figure}

To upgrade the classical solution by the quantum-induced scale dependence, we identify $k$ with  the curvature scale to some appropriate power.
For the classical Schwarzschild case, the Kretschmann scalar is
\be
K = \frac{48G_0^2\, M^2}{r^6}.
\ee
On dimensional grounds, we choose
\be
k^2= \alpha K^{\frac{1}{2}},
\ee
with a dimensionless $\alpha$ of order one. It is convenient to choose $\alpha= 48^{-1/2}$ (such choices can be reabsorbed in the free parameter $\gamma$), such that
\be
k^2=\frac{M G_0}{r^3}.
\ee
This results in the following ``upgraded" line element
\be
ds^2= -f(r)dt^2+f(r)^{-1}dr^2 +r^2 d\Omega^2,\label{eq:dsSchw}
\ee
with 
\be
f(r) = 1-2M\frac{G_N(r)}{r} = 1-\frac{2M}{r\, M_{\rm Pl}^2}\frac{1}{1+\gamma \left(\frac{M}{M_{\rm pl}^4\, r^3}\right)},
\label{eq:f_asimproved}
\ee
where we have set $G_0=M_{\rm Pl}^{-2}$ in the second step, i.e.~we work in natural units. Due to the scale-dependence of $G_N$, working in units with $G_0=1$, as typically done in GR, is not suited to our setting.
In natural units the gravitational radius is given by $r_g = M/M_{\rm Pl}^2$. We measure all radial distances in units of $r_g$.\\
Further, we absorb a factor of $M^2$ defining 
\be
\tilde{\gamma}=\frac{\gamma\,M_{\rm Pl}^2}{M^2}. 
\ee
By working with $\tilde{\gamma}$, all our results apply to arbitrary ratios of $M/M_{\rm Pl}$. Here, we already anticipate that in a more general new-physics setting, one possibly needs to explore significantly larger $\gamma$.\\
With these definitions, the function $f(r)$ in the line element simplifies to
\be
f(r) = 1-\frac{2\,r^2/r_g^2}{r^3/r_g^3 + \tilde{\gamma}}\;.
\label{eq:f_asimproved_gammaTilde}
\ee
Note that the corresponding spacetime is regular and retains a horizon whose
location is given by
\begin{align}
r_H/r_g =&
\frac{1}{6}\Big[4
	+\frac{8\times2^{1/3}}{\left(16-27 \tilde{\gamma }+3 \sqrt{3 \tilde{\gamma }^2 (-32+27\tilde{\gamma})}\right)^{1/3}}
+2^{2/3}\left(16-27 \tilde{\gamma }+3 \sqrt{3 \tilde{\gamma }^2 (-32+27\tilde{\gamma})}\right)^{1/3}\Big]\;.\label{eq:rhSchw}
\end{align}
In the limit $\tilde{\gamma} \rightarrow 0$, where the modifications are switched off, $r_H$ approaches the value of the Schwarzschild solution, $r_H \rightarrow 2\,r_g$, as expected. 
In particular, this limit is approached for $M \gg M_{\rm Pl}$ at fixed $\gamma$.
The horizon shrinks as $\tilde{\gamma}$ is increased, cf.~figure~\ref{fig:rgimprovedhorizon}.  For all $\tilde{\gamma}$, the quantum-gravity inspired black-hole solution has a smaller horizon than its classical counterpart. The difference becomes more pronounced as $\gamma$ is increased, i.e.~the onset of quantum-gravity effects is pushed towards lower curvature scales. 

The RG-improved line element takes a de-Sitter-like form for small $r$, implying a second zero of $f(r)$ below the outer horizon. As $\tilde{\gamma}$ is increased, the two zeros of $f(r)$ move toward each other and annihilate at a critical $\tilde{\gamma}_{\rm crit} = 32/27$, such that the remaining object no longer features a horizon \cite{br00}. 
For the remainder of this paper, we focus on smaller values of $\tilde{\gamma}<\tilde{\gamma}_{\rm crit}$, where no such drastic modification of the causal structure of the spacetime occurs. 
In appendix \ref{app:kink_appendix} we discuss additional features of the shadow in the axisymmetric spacetime, which occur just before the horizon disappears. 
For even larger spin or larger $\tilde{\gamma}$ the horizon disappears and leaves behind a horizonless compact object. Understanding simulated EHT-images for such horizonless spacetimes is an intriguing future question that is beyond the scope of this work.

\subsection{Improved Kerr spacetime}\label{sec:kerr}
Astrophysically, black holes (supermassive or stellar-mass ones) are
  formed in non-spherical systems, either from rotating gas clouds,
  binary mergers, or violently exploding massive stars. Observationally,
  there is growing evidence for non-zero black-hole spin as indicated,
  e.g.~by measurements of the $Fe$ $K\alpha$ emission line profiles
  \cite{2003PhR...377..389R,2014SSRv..183..277R}. Independently, the recent LIGO detections
  \cite{Abbott:2016blz,LIGOScientific:2018jsj} of GWs strongly favour black holes with non-zero
  spin. More generally, even if two merging black holes had vanishing
  initial spins, they would form a Kerr black hole of spin $a\sim 0.69\, r_g$
  \cite{2009ASSL..359..305P}. Moreover, one of the leading models
  to explain jet formation, the Blandford-Znajek mechanism
  \cite{Blandford:1977ds} demands non-zero black-hole spin and fuels
  the jets by energy extraction from the ergosphere via magnetic fields.
  Finally, the EHT observations in conjunction with the jet-power estimates  provide indications for a non-zero spin of M87* \cite{paper5}.

This strongly motivates us to take a step beyond spherical symmetry, and consider the asymptotic-safety inspired upgrade of the Kerr spacetime. The classical line element in Boyer-Lindquist coordinates is given by 
\bea
ds^2&=& 
-\frac{\Delta_r - a^2\,\sin(\theta)^2}{\rho^2}\,dt^2
+\frac{\rho^2}{\Delta_r}\,dr^2
+\rho^2\,d\theta^2
+\frac{(a^2+r^2)^2 - a^2\,\Delta_r\,\sin(\theta)^2}{\rho^2}\sin(\theta)^2\,d\phi^2
\nonumber\\&{}&
-\frac{2(a^2+r^2-\Delta_r)}{\rho^2}a\,\sin(\theta)^2\,dt\,d\phi\;.
\label{eq:dsclassKerr}
\eea
Herein, the specific angular momentum is given by $a= J/M$ and lies between 0 and  $r_g$. Further, we use
\bea
\rho^2&=&r^2+a^2\cos^2\theta,\\
\Delta_r&=&r^2+a^2-2 G_N\, M\, r.
\eea
The corresponding classical Kretschmann scalar which forms the basis for our scale identification takes the form
\bea
K(r,\theta, a)&=&\frac{48 G_0^2 M^2}{(r^2+a^2\cos(\theta)^2)^6} \Bigl(r^6-15 r^4 a^2 \cos(\theta)^2+15 r^2 a^4\cos(\theta)^4-a^6\cos(\theta)^6 \Bigr).\label{eq:KKerr}
\eea
 To date, all RG improvements of the Kerr solution have used a spherically symmetric scale identification  \cite{Reuter:2006rg,Reuter:2010xb,Litim:2013gga,Pawlowski:2018swz}. Yet the curvature actually only diverges in the equatorial plane, and accordingly grows in a $\theta$-dependent fashion, as one approaches the singularity. This suggests that singularity-resolving quantum effects should be strongest in the equatorial plane. Our curvature-informed RG improvement will therefore exhibit an angular dependence, in addition to the radial dependence, as in the Schwarzschild case.  Intuitively speaking, ingoing geodesics at different angles thus differ in their sensitivity to quantum effects.
 
 \begin{figure}
 \centering
\includegraphics[width=0.485\linewidth]{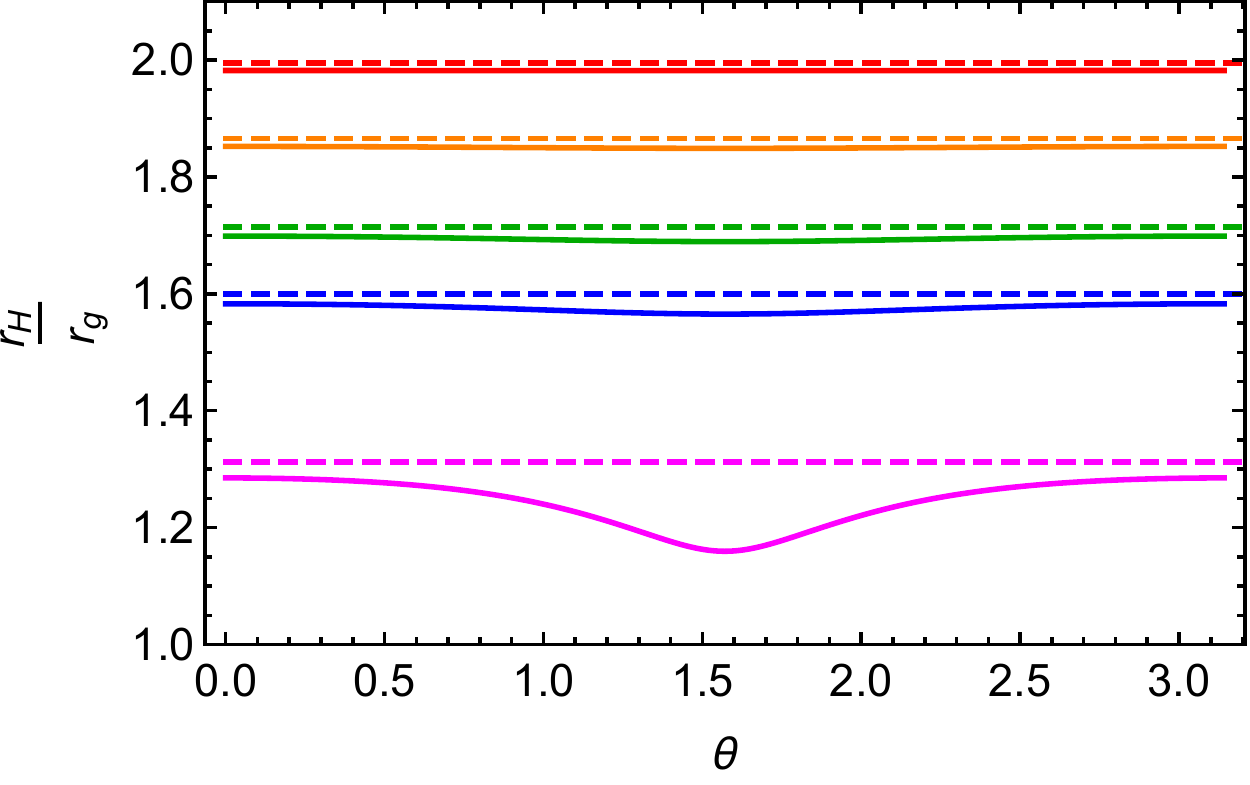}
\hfill
\includegraphics[width=0.485\linewidth]{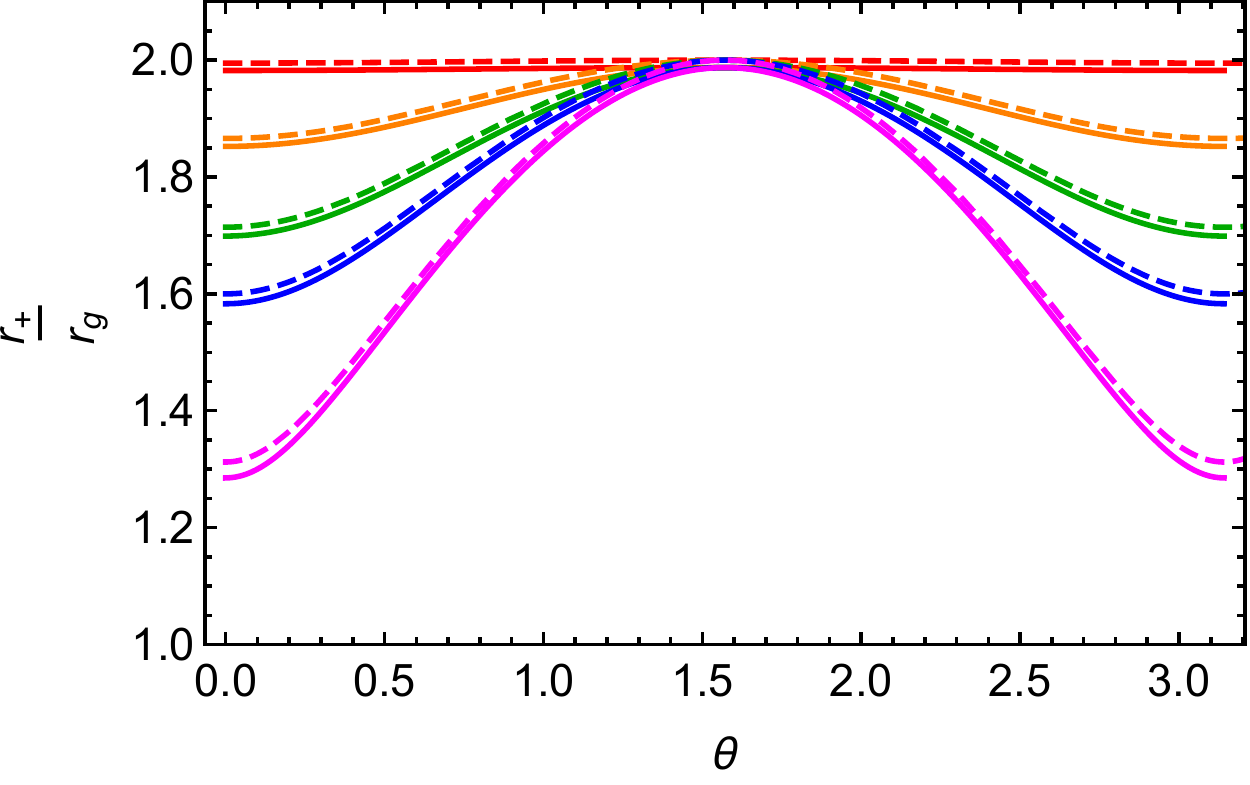}
\caption{\label{fig:Kerrhorizon}
Left panel: We show the classical location of the outer horizon (dashed lines) and the quantum-improved case (continuous lines)  for $\gamma=0.05$ from top to bottom for $a=0.\, r_g$ (red), $a=0.5\,r_g$ (orange), $a=0.7\,r_g$ (green), $a=0.8\,r_g$ (blue) and $a=0.95\,r_g$ (magenta). Right panel: Location of the outer boundary of the ergoregion in the same cases.}
\end{figure}

Following a dimensional argument, we would again identify
\be
k^2=\alpha\, K(r,\theta, a)^{\frac{1}{2}},\label{eq:ksqKerr}
\ee
and choose $\alpha =48^{-1/2}$ for simplicity. However, it is crucial to note that for specific values of $a$ and $\theta$, the Kretschmann scalar $K$ becomes negative, even outside the horizon.
This would result in an imaginary, and therefore unphysical $k^2$. On the other hand, one would expect only the magnitude of $K$ to set the relevant scale. 
This motivates our RG-improvement of the form
\be
k^2 = \frac{G_0 M}{(r^2+a^2\cos(\theta)^2)^3}r^3,\label{eq:Kerr_RG_impr_impr}
\ee
which corresponds to an identification with the main fraction in the Kretschmann scalar Eq.~\eqref{eq:KKerr}, neglecting the $\theta$-dependent polynomial.
The RG-improved metric outside the horizon takes the form Eq.~\eqref{eq:dsclassKerr} with $G_N(k^2)$ and $k^2$ as in Eq.~\eqref{eq:Kerr_RG_impr_impr}.  As a consequence of the RG improvement, the black-hole spacetime is regular.
The RG-improved axisymmetric spacetime reduces to its spherically symmetric counterpart for $a\rightarrow 0$, as it should.

As in the Schwarzschild case, the horizon lies at lower values of $r$. Additionally, it exhibits a $\theta$-dependence that becomes more pronounced as the deviation from spherical symmetry grows, i.e.~as $a$ is increased, cf.~upper panel of figure~\ref{fig:Kerrhorizon}. Due to the fall-off of the RG-improvement effect with radial distance, the modification to the boundary of the ergoregion is smaller, cf.~lower panel of figure~\ref{fig:Kerrhorizon}. As a consequence, the ergoregion for the RG improved black-hole spacetime is slightly larger in volume than in the classical case.

\section{Methods}\label{sec:methods}
Generating an image of the horizon as seen by a distant observer requires tracing light rays through the corresponding black-hole geometry. Pioneering first approaches evolved geodesics forward in time from source to observer and considered radiative transfer from optically thick, geometrically thin accretion disks \cite{Bardeen1973, Cunningham:1975zz}. \cite{Luminet1979} first traced geodesics backwards in time from observer to source, which, in the case of homogeneous and non-localised sources, is far more efficient for a typically very localised observer. We will use similar techniques here. More recently, efficient methods \cite{Viergutz1993, Beckwith:2004ae} for ray tracing involving both a localised source and localised observer have also been developed.

\subsection{Geodesic equation}
In a given spacetime the trajectories of light rays are governed by the null geodesic equation 
\begin{align}
	\label{eq:geodesicEq_null}
	\frac{d^2x^\rho}{d\lambda^2} = -\Gamma^\rho_{\mu\nu}\frac{dx^\mu}{d\lambda}\frac{dx^\nu}{d\lambda}\;,
\end{align}
where $x^\rho$ is the
position of the photon; $\lambda$ is an affine parameter parameterising the photon's world line; and $\Gamma^\rho_{\mu\nu}$ is the  metric-compatible Christoffel connection. We implement the geodesic equation \eqref{eq:geodesicEq_null} as eight coupled first-order ordinary differential equations
\begin{align}
	\frac{dx^\rho}{d\lambda} = k^\rho\;,\quad
	\frac{dk^\rho}{d\lambda} = -\Gamma^\rho_{\mu\nu}\,k^\mu\,k^\nu\;.
\end{align}
For the purpose of this paper, $\Gamma^\rho_{\mu\nu}$ are given in analytical form and since computation time is not critical, we use the native numerical integration techniques available in Mathematica \cite{Mathematica}. 

\subsection{Camera setup \& image}
We position a distant virtual camera far away from the black hole where the geometry is well-approximated by flat spacetime. We will optimise this distance with respect to precision and computation time in the following section. The coordinates of the origin of the image plane in Boyer-Lindquist coordinates 
are
given by $(r_\text{cam},i ,\phi_\text{cam})$. The image plane itself is spanned by two Cartesian coordinates $(x,y)$. Each point 
in
this image plane can be expressed in Boyer-Lindquist coordinates $(r,\theta,\phi)$ of the black-hole spacetime by the following transformation
\begin{align}
    \label{eq:screenCoordsInBL}
    r^2 = \sigma + \sqrt{(\sigma^2 + a^2\,Z^2)^{1/2}}
    \;,\;
    \cos{\theta} = Z/r
    \;,\;
    \tan{\phi} = Y/X\;,
\end{align}
where $\sigma = \left(X^2+Y^2+Z^2-a^2\right)/2$. Here, $(X,Y,Z)$ are Cartesian coordinates centred around the black hole. They are in turn related to the image coordinates $(x,y)$ by
\begin{align}
    X &= \mathcal{D}\,\cos\phi_\text{cam} - x\,\sin\phi_\text{cam}\;,
    \\
    Y &= \mathcal{D}\,\sin\phi_\text{cam} + x\,\cos\phi_\text{cam}\;,
    \\
    Z &= r_\text{cam}\,\cos(i)  + y\,\sin(i) \;,
\end{align}
where $\mathcal{D} = \sin(i)\sqrt{r^2_\text{cam} + a^2} - y\,\cos(i)$.
\\
All the light rays are initialised perpendicular to the screen in which case their initial momentum vector can be calculated by differentiating Eqs.~\eqref{eq:screenCoordsInBL}. 
\\
We parameterise the shadow boundary in the $x-y$ image plane by its radial distance from the origin $\rho(\psi)$ in the image plane as a function of the angle $\psi$ between the $x$-Axis and the radial vector, cf.~figure~\ref{fig:kerr_a}.
The resulting shadow boundary is determined by bisecting nested radial intervals for each $\psi$: Depending on whether the light ray crosses the horizon (and metric components in Boyer-Lindquist coordinates diverge), or escapes to large radii, the outer or inner interval is chosen for the next iteration.
\\
To obtain the intensity distribution generated by a homogeneous background source, we employ the affine-parameter emissivity approximation \cite{Dexter:2009fg}. We normalise the resulting intensity to the image point with the smallest affine parameter.

\subsection{Error control}
We use the deflection angle $\vartheta$ in the equatorial plane as a benchmark value for our error control. In classical Kerr spacetime this angle can be obtained from an analytical form with arbitrary precision \cite{Iyer:2009wa}.
In controlling the initial-data error, the discretisation error (due to a finite stepsize), and the computational errors due to finite numerical precision, we rely on standardised and well-known error control of the native ODE-solver, i.e.~\cite{Mathematica}.
\\
Additionally, there is an error due to the finite radial camera distance. At large distance $r_\text{cam}\gg r_g$ all investigated black-hole metrics converge to flat space. In this regime, the dependence of the deflection angle $\vartheta$ on the radial distance is therefore expected to obey the functional form
\begin{align}
	\label{eq:error_deflectionAngle}
    \vartheta_\text{fit}(r_\text{cam}) = \vartheta_0 - b/r_\text{cam}\;.
\end{align}
We fit this function to a series of data points obtained at increasing values of $r_\text{cam}$ to determine the parameters $\vartheta_0$ and $b$. Specifying a chosen maximal error $\Delta\vartheta = \vartheta_\text{fit}(r_\text{cam}) - \vartheta_0$ determines the required radial distance $r_\text{cam}(\Delta\vartheta)$.
In App.~\ref{app:error-control-kerr}, we apply this procedure to fit the exact known result for Kerr spacetime. The exact form for $\theta$ in the equatorial plane allows us to also benchmark the required numerical precision. In App.~\ref{app:error-control-kerr}, we also demonstrate the radial distance error control for an explicit deflection angle in the equatorial plane of the RG-improved spacetime.
\\
Kerr-like spacetimes exhibit three constants of motion: the energy $E$, the angular momentum along the black-hole rotation axis $L_Z$, and the celebrated Carter constant $\mathcal{Q}$ \cite{Carter:1968rr}.
We use the conservation of energy and the angular momentum as independent checks of our numerical error.

\section{Results}\label{sec:results}
We have motivated the RG-improved metrics for the spherically symmetric and axisymmetric case from asymptotically safe gravity. Yet, quantum gravity is just one specific candidate for new physics leading to singularity resolution. From a more agnostic point of view, the study of singularity-free, black-hole like geometries should proceed from a vantage point independent of one specific theory. In the present context, this implies that the parameter $\gamma$ is not required to be close to 1, as the scale where modifications set in need not be tied to the Planck scale. Thus, it is of interest to constrain the resulting, large parameter space which can most efficiently be done by strong-field observations. 
Working with $\tilde{\gamma}$ instead of the original parameter $\gamma$ allows us to investigate the modifications to shape and size of the shadow in a black-hole mass-independent fashion, as $\tilde{\gamma}$ scales appropriately with $M$, whereas the magnitude of the effect decreases  at fixed $\gamma$ as $M$ is increased.

\subsection{Spherically symmetric black-hole spacetimes and their shadows}
\begin{figure}
\centering
\includegraphics[width=0.6\linewidth]{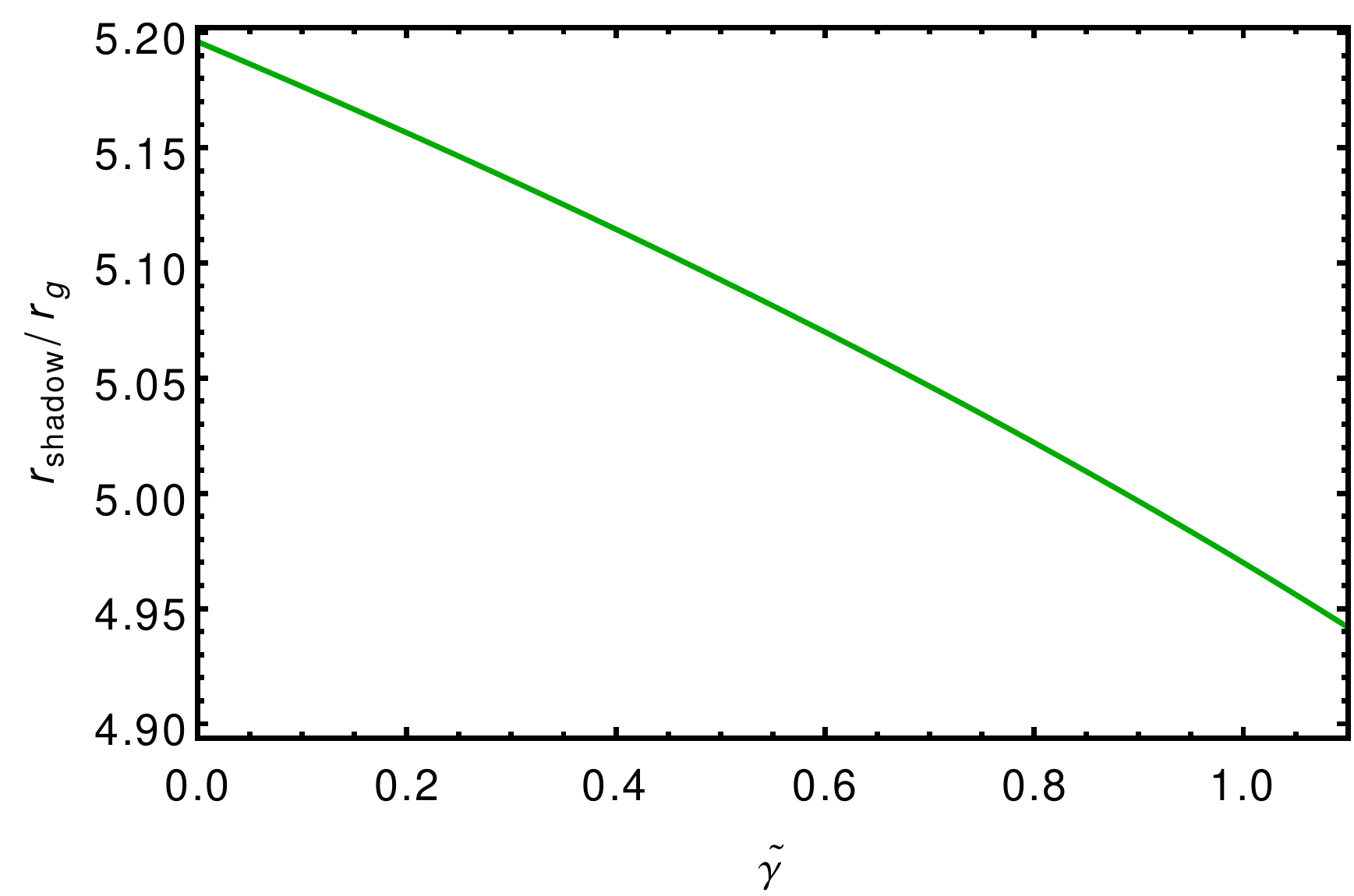}
\caption{\label{fig:rshadowgamma}Coordinate location of the shadow for the spherically symmetric spacetime as a function of $\tilde{\gamma}$. The classical case is $\tilde{\gamma}=0$.}
\end{figure}

We first focus on the spherically symmetric case.
We observe that the size of the shadow decreases as a function of $\tilde{\gamma}$, cf.~figure~\ref{fig:rshadowgamma}. Intuitively, this is to be expected: In our setting, quantum-gravity effects lead to a weakening of the gravitational interaction strength. At a given proper distance to the centre of the black hole, the escape velocity is lowered compared to the classical case, as an observer leaving the black hole in their rocket experiences less gravitational pull in the quantum-improved case.
Therefore, the "point of no return" for any infalling observer lies closer to the centre.
Accordingly, weaker gravity leads to a more compact event horizon.

\begin{figure}
\centering
\includegraphics[width=0.6\linewidth]{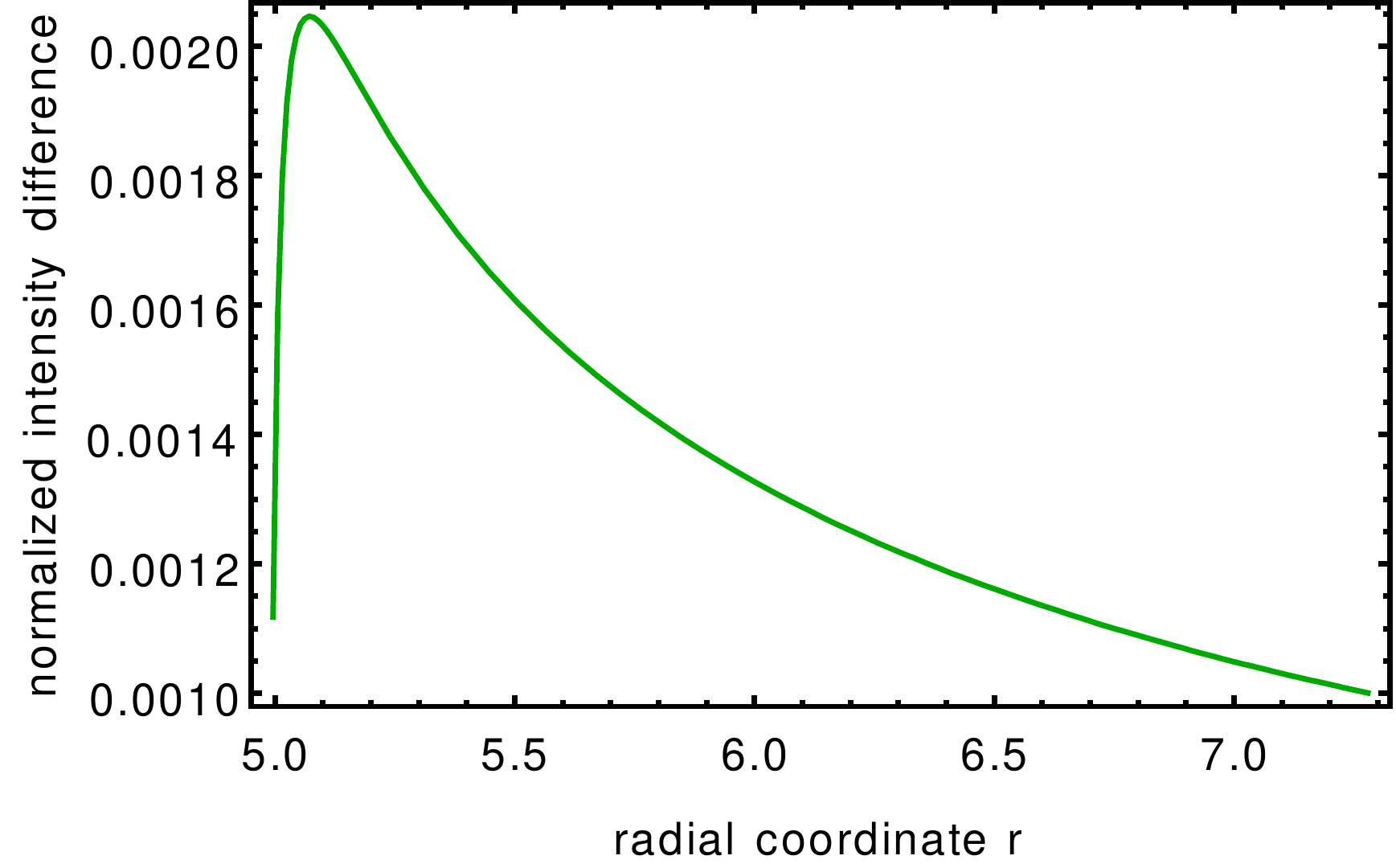}
\caption{\label{fig:intensitycomp} Relative intensity difference (within the approximate emissivity scheme) as a function of $r$, in units of $r_g$,  between the RG improved black-hole spacetime and classical Schwarzschild spacetime with $M'<M$, such that both feature the same shadow size. }
\end{figure}

The modifications of the spacetime resemble that of a classical Schwarzschild black hole of lower mass. 
This begs the question whether there is a degeneracy between the classical Schwarzschild solution of mass $M$ and the RG-improved solution of mass $M'<M$. Indeed, this is the case if the size of the shadow is the only observable that is accessible. Yet, a comparison of the intensity distributions in the two cases already demonstrates that there are additional differences, cf.~figure~\ref{fig:intensitycomp}.

The modifications due to RG improvement are \emph{not} a simple rescaling of a classical Schwarzschild solution in the form $M \rightarrow M'$.  Specifically, one can rewrite
the RG-improved metric in the form of the classical Schwarzschild metric
with an \emph{effective} mass that depends on the radial distance,
obtained by rearranging Eq.~\eqref{eq:f_asimproved},
\be
M_{\rm eff}(r)= \frac{M}{1+\tilde{\gamma}\,\left(\frac{r_g}{r}\right)^3}.\label{eq:Meff}
\ee
At the horizon, the difference between $M_{\rm eff} (r)$ and $M$ is largest, and the modifications fall off with $r^{-3}$. This highlights that the deviation from classical Schwarzschild that we explore is a strong-field effect: while the deviation is at the level of about 10\% close to the horizon (for $\tilde{\gamma} \approx 1$), it falls off rapidly with increasing radius. 

The classical Schwarzschild spacetime and the RG-improved case
can be distinguished via the $r$-dependence in Eq.~\eqref{eq:Meff}.  Specifically, one can in principle extract the effective mass at two
different distances from the black hole. For instance, one measurement can
be extracted from the size of the shadow at distances $\sim M$. The
second could use the Keplerian orbital periods of nearby stars
at distances of $\gtrsim 10^3M$ (e.g.~the pericentre distance of the star
  S2 in the galactic centre \cite{Ghez:2008ms,Ghez:1998ph,Gillessen:2008qv,Meyer:2012hn}). Weighing black
holes using the orbital motion of nearby stars is still possible, even
for spatially unresolved orbital motion, via spectral analysis of
emission line profiles of stars or gas in orbit around a supermassive
black hole, as done for M87 \cite{Gebhardt:2011yw,Walsh:2013uua}. For a
classical Schwarzschild spacetime, these results should agree. In the RG-improved case, the effective mass extracted from the size of the shadow
is smaller than the effective mass extracted at larger radii (for small $\gamma$, i.e.~effects tied to the Planck scale, only marginally so). This makes it evident that the modifications of the spacetime are not degenerate with a classical spherically symmetric spacetime.

 For the case of M87*, the combined statistical and systematic error on the mass measured by the EHT collaboration is roughly 14 \% \cite{paper6}. Stellar-dynamics measurements yield an error of below 10 \% \cite{Gebhardt:2011yw}, while gas dynamics observations have a significantly higher error \cite{Walsh:2013uua} (note that these results are not in agreement with each other). Assuming that the near-horizon and far-horizon measurements should differ by one $\sigma$, these accuracies can be used to constrain $\tilde{\gamma}\lesssim 2$. Converted into $\gamma$, the higher mass of M87* compared to Sgr A* results in $\gamma \lesssim 2 \cdot 10^{95}$. This result highlights that modifications tied to the Planck scale will not be accessible by EHT observations. However, non-quantum, singularity-resolving gravitational physics might very well exist, and EHT observations do provide a way of probing or constraining the corresponding effects. We also highlight that the constraint $\tilde{\gamma}<2$ cannot yet exclude horizon-less objects described by our RG improved metric. Future observations at higher frequencies and/or in polarization data, see \cite{Gold:2016hld} can result in tighter constraints, potentially ruling out horizonless objects in our setting.

Our result motivate a closer look at Sgr A*, given that at fixed $\tilde{\gamma}$, every order of magnitude in the mass corresponds to two orders of magnitude in the constraint on $\gamma$. 
  
For  Sgr A*, weak-field mass measurements come from tracking the orbital motion of stars around Sgr A* \cite{Gillessen:2008qv, 2015MNRAS.447..948C, 2018A&A...618L..10G, 2018A&A...615L..15G, 2019PhRvL.122j1102A, 2019arXiv190405721A}. The most recent result provides an accuracy of $0.3\%$, cf.~\cite{2019arXiv190405721A}.
Furthermore, let us assume that in the future, EHT will provide a measurement of the shadow size of Sgr A* with an accuracy of about 6 \%, see \cite{Johannsen:2015hib}. In our model, these two measurements are expected to differ by one sigma for $\tilde{\gamma}\gtrsim 0.5$. Reinstating the black-hole mass explicitly, this still translates into a constraint of $\gamma \lesssim 10^{89}$.  We emphasize again, that this extraordinarily large number is a consequence of theoretical bias that the scale of new physics should be the Planck scale. Note that the scale of new physics is actually unknown, and therefore any constraint on the parameter space is important to achieve. Of course, if we use a different scale than $M_{\rm Pl}$ as our reference scale, the corresponding expected values for $\gamma$ can increase dramatically, closing the gap to the observational constraints. 

Stronger constraints will be possible if stars on tighter orbits than S2 are found and especially when a pulsar in orbit around a SMBH is found, as studied in \cite{deLaurentis2018}. Further, observations of the stellar dynamics making use of a forthcoming 39-m telescope are expected to provide a measurement of the black-hole mass with an accuracy of 0.1\% \cite{2005ApJ...622..878W}.  Specifically, constraints $\tilde{\gamma} \sim \mathcal{O}(10^{-1})$ would become available if both measurements could reach a sub-percent accuracy. In particular, future observations of Sgr A* will thus probe values below $\tilde{\gamma}_{\rm crit}$ and distinguish horizon-less objects from black-hole spacetimes within our setting.
 \newline

 Let us contrast our results with observations in the weak-field regime. The gravitational parameter space can be spanned by two parameters, measuring the curvature $\xi$ (given by the Kretschmann scalar for Schwarzschild) and the Newtonian potential $\varepsilon$ (linked to an observable, namely the gravitational redshift). The highest curvatures probed in controlled laboratory experiments at low Newtonian potential are actually tests of the Newtonian inverse-square law, see, e.g.~figure~2 in \cite{2015ApJ...802...63B}. Specifically, following this reference, the absence of modifications to the inverse square law down to $56\, \mu m$ reached in \cite{2007PhRvL..98b1101K} corresponds to a probe at $\xi = 6\cdot 10^{-24} m^{-2}$. This should be contrasted with $\xi \approx 10^{-20}m^{-2}$ near the horizon of Sgr A*. Moreover, we highlight the following point: While tests of the inverse-square law proceed at very small $\varepsilon \approx 10^{-33}$ (see \cite{2015ApJ...802...63B}), the EHT is sensitive to sources with $\varepsilon \approx 10^{-1}$.
In the presence of a second scale, set by $\varepsilon$, an RG-improvement $\sim\varepsilon\,\xi$ could also be motivated.
Accordingly, the strong-field regime should provide a qualitatively different test of the effects we propose here.\newline

Beyond the specific model we explore here, we expect that the signature of quantum gravity we find here could be generic for a large class of quantum-gravity theories. For quantum-gravity effects to resolve the classical singularity, an effective weakening of gravity at high curvature scales -- linked to an effective repulsive force from quantum gravity -- is generically expected. This would imply, that quantum-improved black-hole horizons are more compact than their classical counterparts. 
 The shrinking of the horizon (and consequently the shadow) through quantum-gravity effects can also be understood as a consequence of demanding a de-Sitter-like core in which the classical singularity is resolved. Inspecting the line element, we observe that a transition from Schwarzschild behaviour at asymptotically large distances to de-Sitter-like behaviour in the core requires a second zero of the function $f(r)$ in Eq.~\eqref{eq:dsSchw}. This results in a shift of the horizon to smaller values of $r$.
Accordingly, we conjecture that most (if not all) quantum-gravity models will feature a (typically not large enough to be observable) signature of the form we discussed here, namely a more compact black-hole horizon and shadow. This can be reproduced by changing the one free parameter of the classical Schwarzschild solution, the mass, to a smaller value.  The difference between physically different quantum theories of gravity lies in the fall-off of the effective mass, which could be different powers for different theories. Accordingly, a third mass-measurement at an intermediate distance could in principle distinguish these different theories. In practice, such differences are of course tiny, unless the curvature scale at which quantum-gravity effects set in is  rather low.  Remaining more agnostic about the type of new physics that leads to singularity resolution, the effective description in terms of a repulsive force (decreased black-hole mass) remains well-motivated, with $\tilde{\gamma}\sim\mathcal{O}(1)$ not excluded. This calls for dedicated efforts to understand the dynamics of infalling and radiating matter in such spacetimes, in order to bridge the gap to actual EHT-images.

\subsection{Axisymmetric black-hole spacetimes and their shadows}

\begin{figure}
\centering
\includegraphics[width=0.7\linewidth]{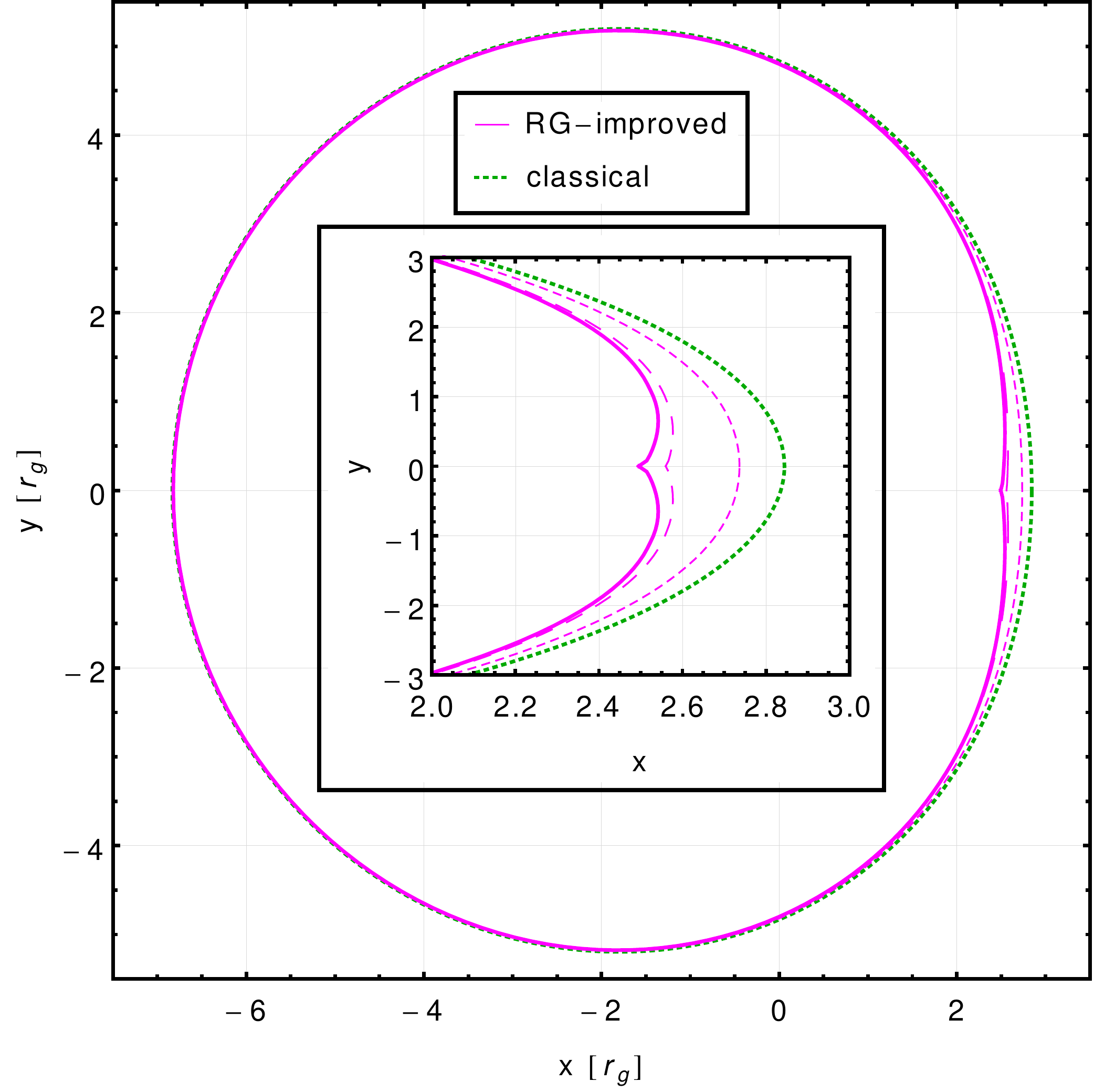}
\caption{\label{fig:kerr_a09_gamma(0and01)}
Classical Kerr shadow (green dotted) for $a=0.9\,r_g$ and RG-improved case (magenta) for $a=0.9\, r_g$. Growing values for $\tilde{\gamma}=(0.05,\,0.1,\,0.11)$ are indicated by smaller to larger dashing, respectively.}
\end{figure}

\begin{figure}
\centering
\includegraphics[width=0.7\linewidth]{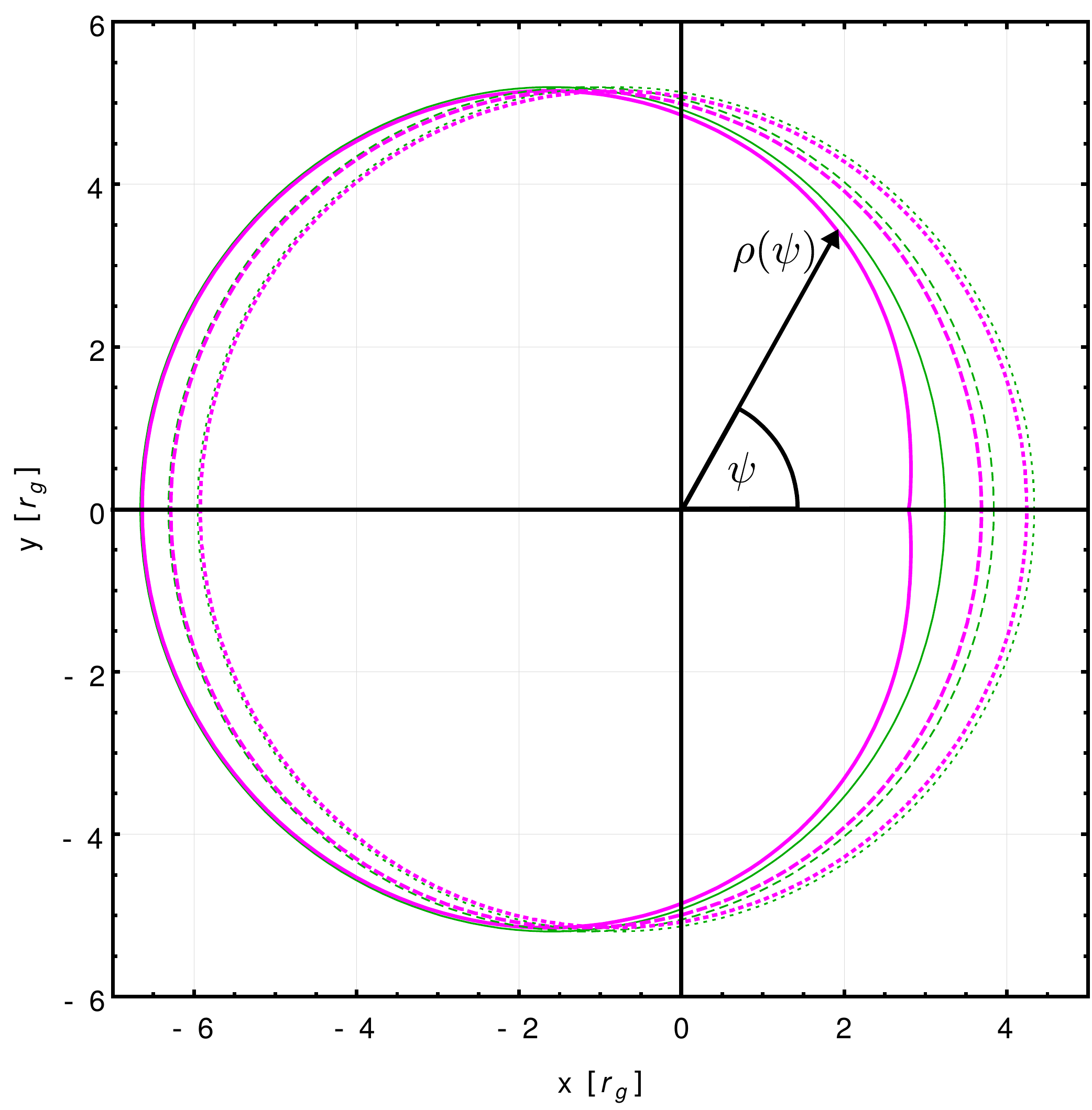}
\caption{\label{fig:kerr_a}
Classical (thin green) and RG-improved (thick magenta) shadows for various spin-parameters $a=0.4\, r_g$ (dotted), $a=0.6\, r_g$ (dashed), and $a=0.8\, r_g$ (continuous). In all cases $\tilde{\gamma} = 0.25$.}
\end{figure}

At leading order, the asymptotic-safety inspired effect is the same in the spherically symmetric and axisymmetric case, and consists in a reduced size of the shadow compared to the classical case, see figure~\ref{fig:kerr_a09_gamma(0and01)}. As a consequence of frame-dragging, rays that end up in the  prograde (right) side of the image pass significantly closer to the horizon than those in the retrograde (left) side of the image. Accordingly, they pass through a regime of larger curvature, where the differences of the RG-improved and the classical spacetime are larger. Therefore, the deviation of the RG-improved shadow from the classical shadow is larger in the prograde (right) side of the image. Furthermore, this effect grows with increasing spin, see figure~\ref{fig:kerr_a}. 

As discussed in Sec.~\ref{sec:kerr}, the black-hole horizon is no longer spherically symmetric, but instead axisymmetric,  in our model. The departure from spherical symmetry is a consequence of the fact that the curvature increases fastest in the equatorial plane, leading to the most pronounced shrinking of the horizon in this plane. This feature also affects the shadow, and can even generate a dent in the shadow if the effects are strong enough, cf.~inset of figure~\ref{fig:kerr_a09_gamma(0and01)}.

\begin{figure}
\centering
\includegraphics[width=0.7\linewidth]{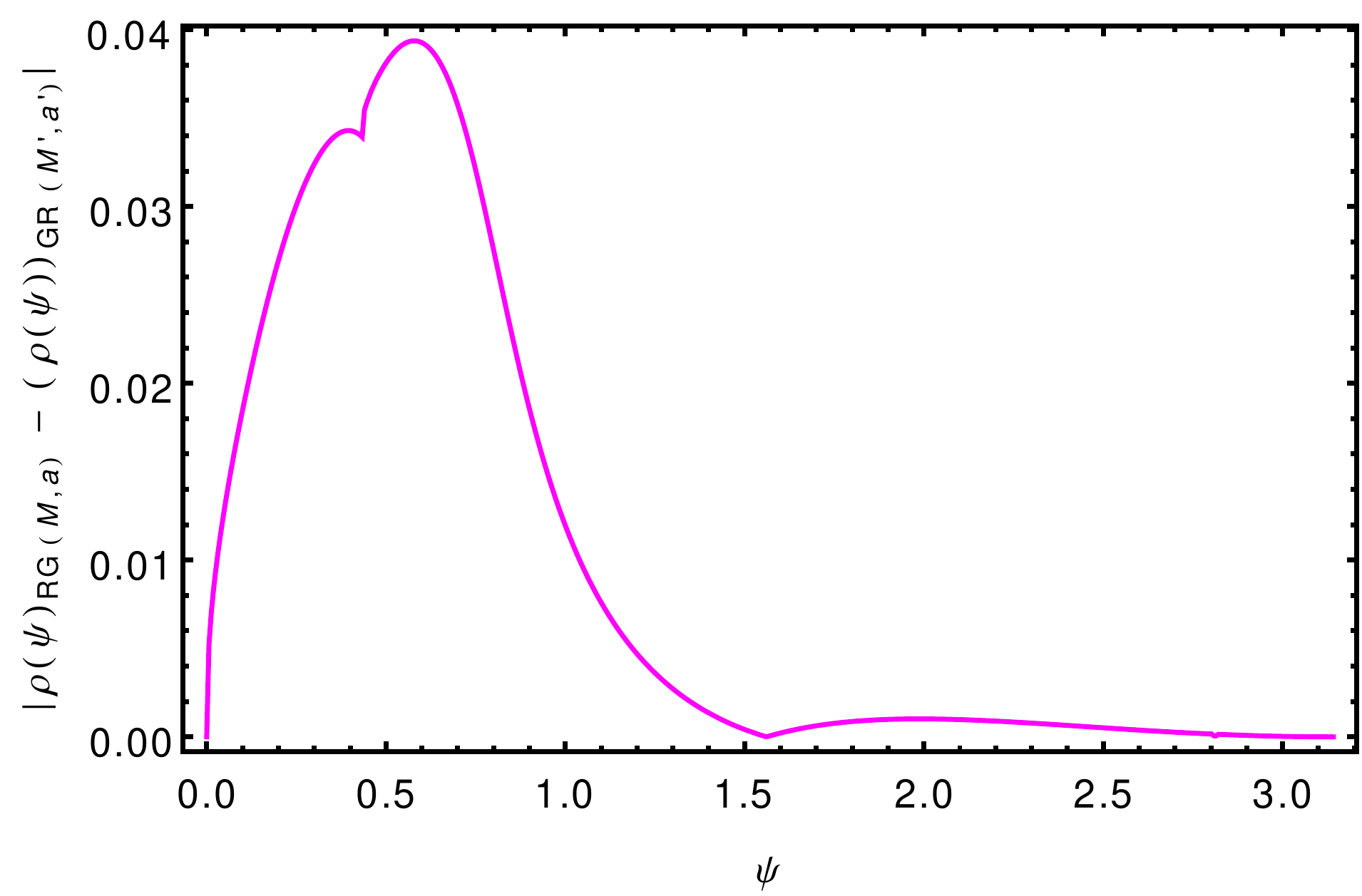}
\caption{\label{fig:radialdifference}
Difference in radial coordinate of the parameterised shadow boundary $\rho(\psi)$ between the RG-improved shadow (for $M=1$, $a=0.99\, r_g$, and $\tilde{\gamma} = 0.008$) and the classical shadow at different mass and spin ($M=1.001435$, $a=0.999209\, r_g$). The different mass and spin are chosen such as to result in degenerate points on the major axes at $\psi=0,\,\pi/2,\,\pi$. Every other point on the shadow boundary lifts this degeneracy.
}
\end{figure}

In contrast to the spherically symmetric case, the shadow shape of the RG-improved black-hole spacetime is no longer degenerate with a classical shadow shape. Adjusting the two parameters $M$ and $a$ in the classical solution in order to match the two major axes of the RG-improved shadow results in differences at every other point in the image, cf.~figure~\ref{fig:radialdifference}. These differences are small and therefore difficult to spatially resolve. Yet, we also point out that comparing the mass at two different radial distances (e.g.~extracted from the shadow, and from Keplerian orbits) still allows to distinguish the classical and the RG-improved spacetimes, just as in the spherically symmetric case, at least for sufficiently large $\tilde{\gamma}$.

\begin{figure}
\centering
\includegraphics[width=0.7\linewidth]{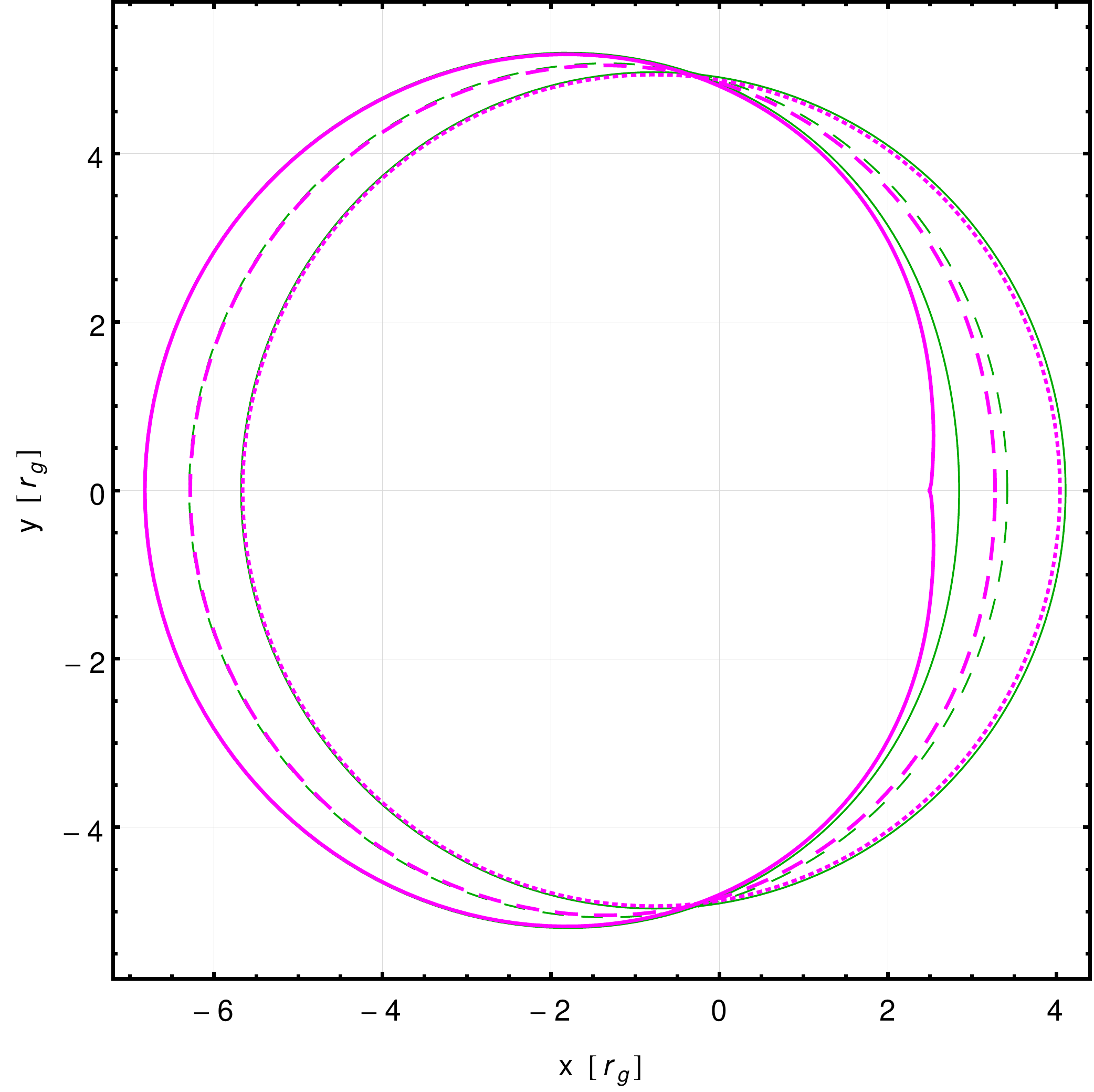}
\caption{\label{fig:kerr_theta_cam}
We show the RG improved case (magenta, thick lines) and the classical shadow (green, thin lines) for $a=0.9\, r_g$ and $\tilde{\gamma}=0.11$. The angle between the spin axis and the observer is $i=\pi/8$ (dotted lines), $i=\pi/4$ (dashed line) and $i=\pi/2$ (continuous line).}
\end{figure}

Next, we explore the image as a function of the inclination i, i.e. the angle between the black hole’s spin axis and the line of sight, see figure \ref{fig:kerr_theta_cam}. As one decreases $i$, the image approaches a circle (just as in the classical Kerr case), and therefore the modifications characteristic of the axisymmetric case become less pronounced.

\begin{figure}
\centering
\includegraphics[width=0.495\linewidth]{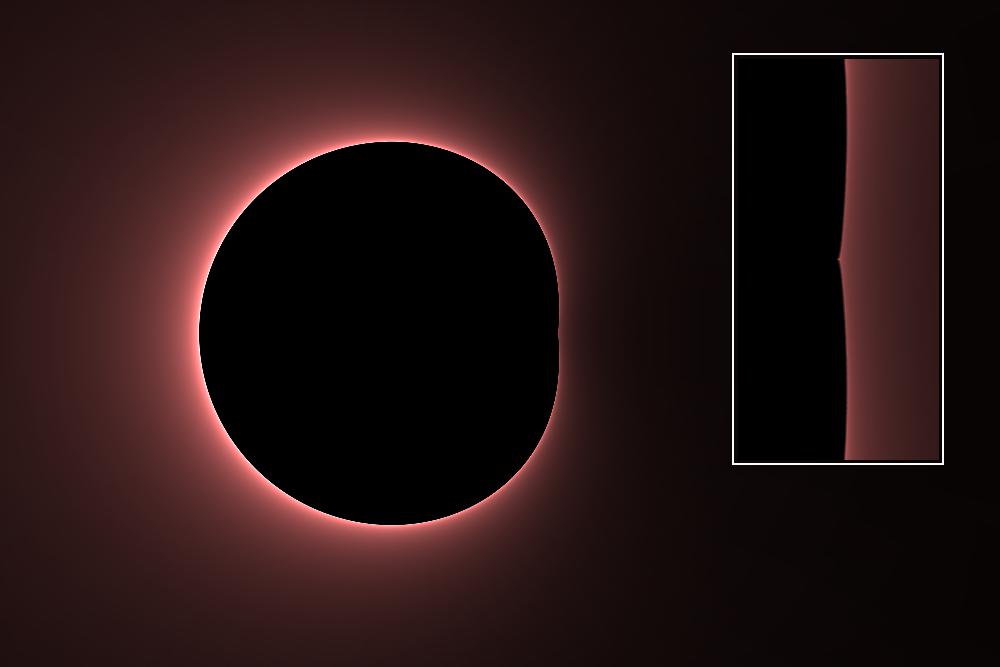}
\hfill
\includegraphics[width=0.495\linewidth]{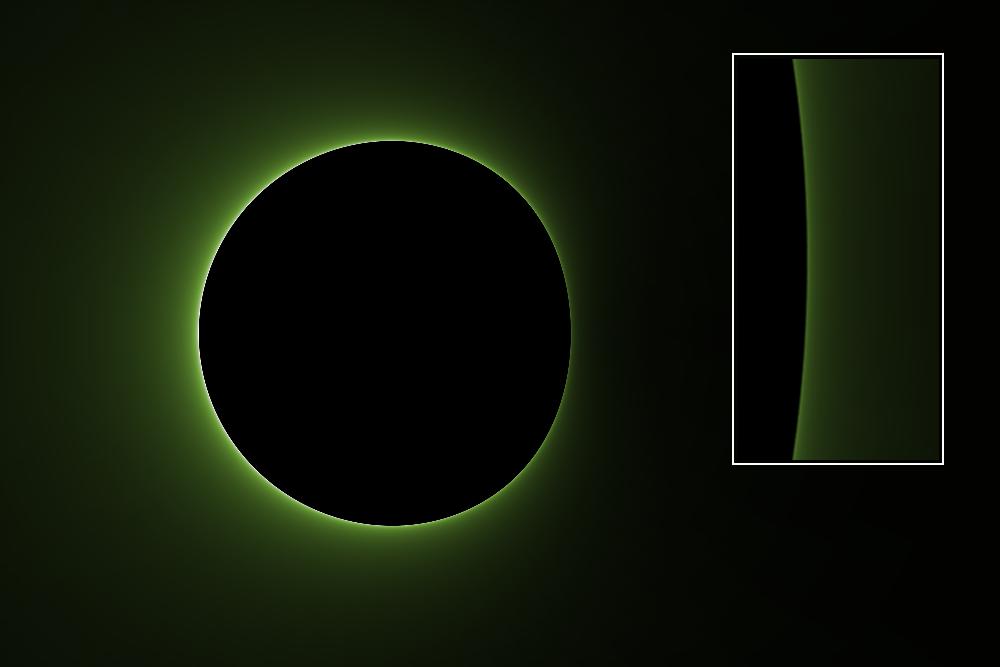}
\caption{\label{fig:kerrintensity}
We show the intensity image extracted from the affine-parameter emissivity for an RG-improved (left panel) and classical (right panel), axisymmetric black-hole spacetime with $a = 0.9\, r_g$ and $\tilde{\gamma}=0.11$. The inset shows a zoom into the prograde (right) side of the image, showing the characteristic dent-like feature in the RG-improved image.}
\end{figure}

Overall, the new features in the shadow are most pronounced for fast-spinning black holes viewed from within the equatorial plane, see figure~\ref{fig:kerrintensity}. If effects are restricted to the Planck scale, they of course remain too tiny to be detectable -- even though modifications are actually always present. We stress that $\gamma$ can alternatively be viewed as a parameterisation of new, singularity-resolving physics which need not be quantum gravity. In that case, $\gamma$ is not tied to the Planck scale, and could be significantly larger. \newline

Apart from asymptotic safety, we argue that the effects we describe here could be universal consequences of a large class of  quantum-gravity theories. Just as in the Schwarzschild case, one might expect singularity resolution to be tied to an effective weakening of gravity at high curvature scales. Accordingly, the modifications of the black-hole spacetime decrease in magnitude with increasing distance from the horizon. For axisymmetric spacetimes, this has important consequences due to frame-dragging: Because of frame-dragging, 
rays ending up in the prograde (right) side of the picture probe the spacetime much closer to the horizon than in the  retrograde (left) side, and are much more affected by the modifications. 
Accordingly, the shadow is more strongly modified on one side only. Thus, a first rather generic consequence of quantum gravity in spinning spacetimes is that stronger effects are expected on the prograde side. Next, we argue that a dent-like feature in the equatorial plane of the image could also be generic: As the curvature in the classical Kerr spacetime becomes singular only in the equatorial plane, the curvature at fixed radial distance is largest in that plane, resulting in stronger quantum-gravity effects via Eq.~\ref{eq:Kerr_RG_impr_impr}. These render the black-hole horizon more compact. Thus, the shadow shrinks more on the prograde side of the picture, and most strongly for $\psi=0$, compared to the classical case. This is expected to result in a dent-like feature, the extent of which is of course subject to non-universal properties of a given quantum-gravity theory. While these effects are always present, they remain too tiny to be detectable for astrophysical black holes, if the scale of the effect is tied to the Planck scale. An earlier onset of quantum-gravity effects for black holes than expected based on simple power-counting might change this.

\section{Discussion}\label{sec:conclusions}
The groundbreaking observations of the Event Horizon Telescope \cite{paper1, paper2, paper3, paper4, paper5, paper6} have provided us with a very first image of a black hole. This image opens up a new window into the strong-field regime of gravity, offering a unique possibility to constrain the fundamental physics of spacetime. To confront viable theories of quantum gravity, and more generally singularity-free models, with the observational data now available, resulting features of black-hole shadows have to be derived. We set out to close this gap for asymptotic-safety inspired black-hole spacetimes. A key feature of asymptotically safe gravity is the dynamical weakening of the Newton coupling beyond a critical high-energy scale which is not necessarily tied to the Planck scale. We upgrade classical black-hole spacetimes by incorporating this scale dependence, assuming that the relevant physical scale is the local curvature scale. Therefore, modifications are largest close to the horizon and thus our work is one example of the general point that the shadow size is a powerful test of gravity \cite{Johannsen:2015hib, Psaltis:2018xkc}.
In constructing the asymptotic-safety inspired spacetime, we for the first time take into account the angular dependence of the local curvature scale in axisymmetric spacetimes, and add to the evidence that the scale-dependence of the Newton coupling could lead to singularity-resolution.

Let us first highlight the qualitative nature of the resulting effects before commenting on their size.
As a consequence of the weakening of gravity at high curvature scales, the upgraded spacetimes we explore feature a more compact horizon. Intuitively, this is a consequence of an effective repulsive force arising due to quantum effects. As our first key result, this leads to a smaller shadow both in the spherically symmetric and axisymmetric case. 
In the former case, the shadow is degenerate with that of a classical black-hole spacetime of smaller mass. 
In fact, the asymptotic-safety inspired upgrade of the spherically symmetric spacetime can be parameterised by a mass function depending on the radial distance.  This mass function approaches the classical mass from below as a function of increasing distance to the horizon.
Therefore, the degeneracy in shadow-size between the classical and upgraded spacetime could be lifted by using a second observable at different radial distance. Specifically, a second mass measurement extracted from Keplerian orbits breaks the degeneracy.

Moreover, there is no degeneracy in spinning spacetimes. Due to frame-dragging, rays arriving at different points in the image plane probe the spacetime at different curvature scales (i.e.~different distances to the horizon). In particular, in the near-extremal black-hole case, a subset of light rays probe the spacetime arbitrarily close to the horizon, and are therefore most sensitive to the increased compactness.
The latter is most pronounced at the equator of the spinning black hole.
Thus, generically, we find that the asymptotic-safety inspired effects put a dent into the shadow close to the equator in the image. Naturally, if quantum-gravity effects are suppressed by the Planck scale, these features are present, but undetectable for astrophysical black holes.\newline

We argue that the two features (the overall shrinking of the shadow and the dent close to the equator) ought to be generic consequences of a large class of quantum-gravity theories. Specifically, we argue that lifting a classical singularity requires an effective repulsive force from quantum gravity. This leads to a more compact black-hole shadow. In the axisymmetric case, we expect this to lead to stronger effects on the flattened side of the shadow, in particular in the equatorial plane. In fact, Loop Quantum Gravity is another example where indications have been found for an effective repulsive character of quantum effects, see, e.g.~\cite{Ashtekar:2006uz}, underlying a scenario for regular black holes \cite{Modesto:2008im,Gambini:2008dy,Gambini:2013ooa,Rovelli:2014cta}. These feature a smaller horizon compared to the classical case, just as in our model. Moreover, a non-commutative structure of spacetime has also been argued to lead to a regular black-hole spacetime with a smaller horizon than in GR \cite{Nicolini:2005vd}. Further, stringy corrections motivate a similar result \cite{Nicolini:2019irw}.
Accordingly, the features we discover here could constitute blueprints for generic quantum-gravity effects.

Within quantum gravity, the most natural scale for the departure from GR to set in, is approximately the Planck scale. Yet, this is based on simple dimensional analysis, and need not hold in a fully dynamical theory. 
Widening our scope, we point out that singularity-resolution is physically necessary, but need not be tied to quantum gravity, and could be a consequence of new classical physics. Any modification that is equivalent to a repulsive force and thus leads to a de-Sitter-like core, will generically lead to a more compact horizon \cite{Dymnikova:1992ux,Dymnikova:1996ux,Hayward:2005gi,Ghosh:2014pba}. 
In this more general case, the typical scale of modifications becomes a free parameter  and our work captures the expected imprints of these modifications on the black-hole shadow. We sketch how observations of the EHT combined with measurements of Keplerian orbits will allow to constrain the free parameter, and motivate an exploration of modified black-hole spacetimes of the form we study here.\newline

To conclude, we highlight that we have computed the shadow for  spacetimes which i) feature a horizon, ii) are regular due to a physical mechanism motivated by quantum gravity and iii) are expected to be compatible with the recent groundbreaking observation of the EHT collaboration for the quantum-gravity region of parameter space. \newline

The work presented here serves as a natural stepping stone for a more
  advanced model that takes into account actual synchrotron emission of
  relativistic electrons in a turbulent plasma around the black hole as
  expected for the two main EHT sources Sgr A* and M87*. The behaviour
  of the plasma can be obtained from General Relativistic
  Magnetohydrodynamical simulations in the upgraded metric, similar in
  spirit to \cite{Mizuno2018} yielding four-velocities, densities,
  temperatures and magnetic field strengths and structures. These
  predictions then allow evaluations of the synchrotron emissivity
  enabling to predict images from such models that will carry imprints
  of the upgraded black-hole spacetime and a direct comparison to
  data taken by the EHT thereby probing or constraining deviations to GR.

\acknowledgments
This work is partially supported by the ERC Synergy Grant ``BlackHoleCam: Imaging the Event Horizon of Black Holes'' (Grant No.~610058).
 A.~E.~and A.~H.~are supported by the DFG under grant no.~Ei-1037/1, and A.~E.~is also partially supported by a visiting fellowship at the Perimeter Institute for Theoretical Physics and by the Danish National Research Foundation under grant DNRF:90. A.~H.~also acknowledges support by the German Academic Scholarship Foundation and thanks CP3-Origins for hospitality during a part of this project.

\begin{appendix}

\section{Error control in Kerr spacetime}
\label{app:error-control-kerr}
\begin{figure}
\hspace{20pt}
\includegraphics[width=0.4\linewidth]{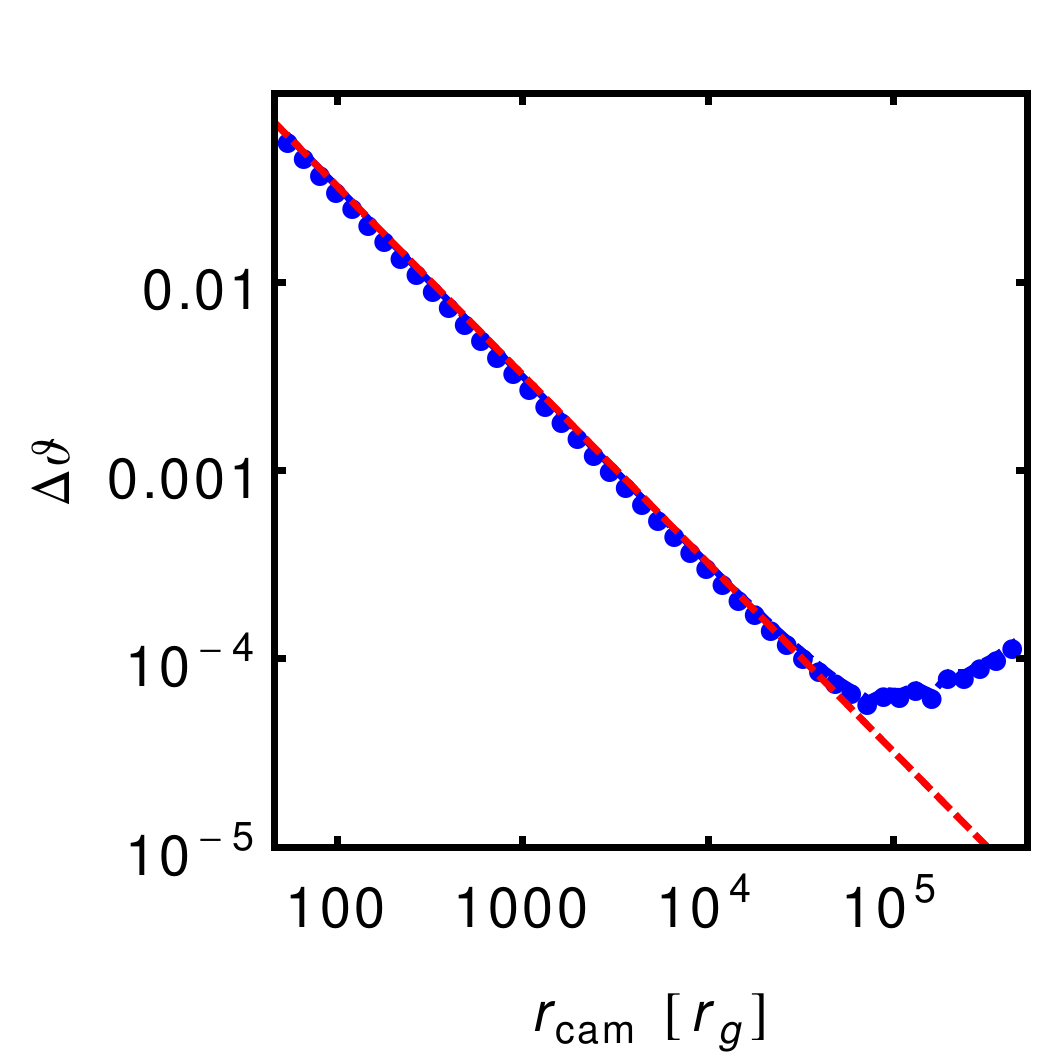}
\hfill
\includegraphics[width=0.4\linewidth]{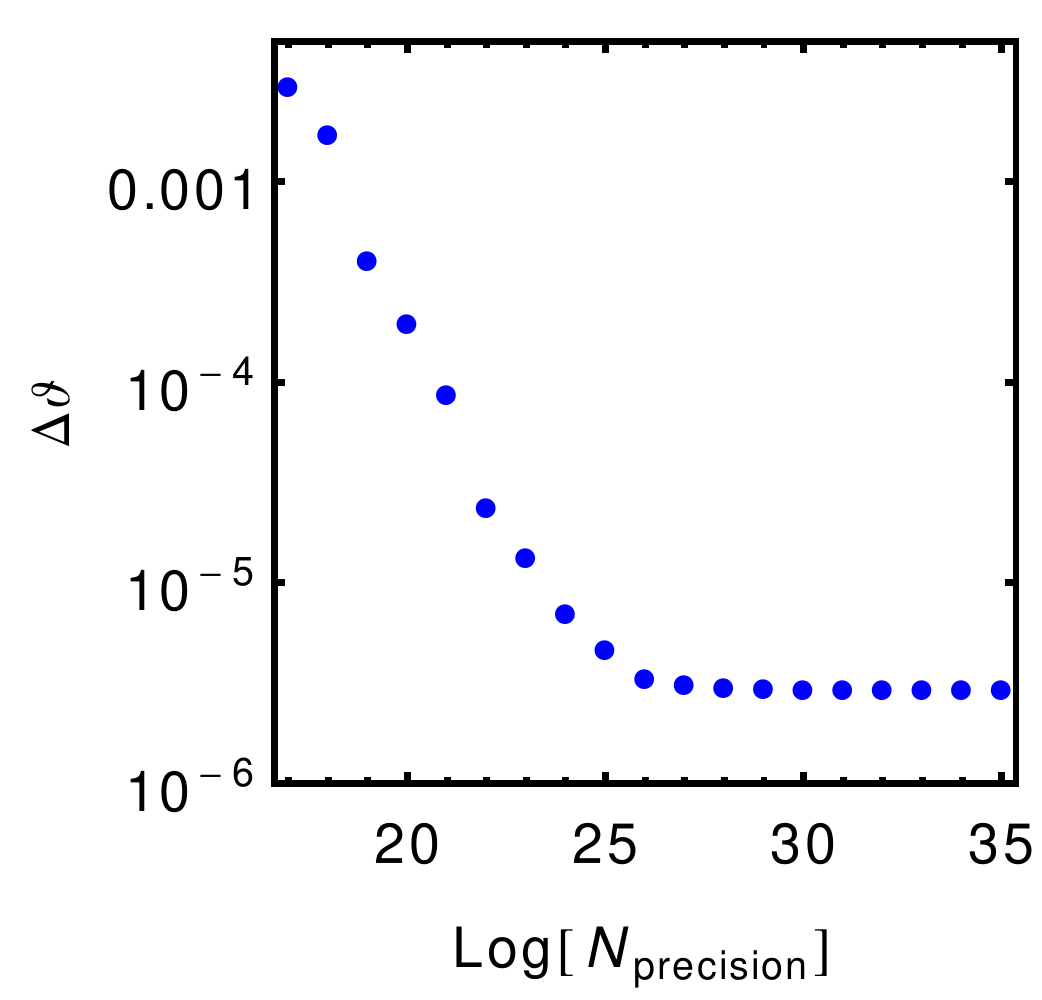}
\hspace{20pt}
\caption{\label{fig:errorKerr}Deflection angle error $\Delta\vartheta = \vartheta_\text{fit}(r_\text{cam}) - \vartheta$, where $\vartheta$ is the exactly calculable deflection angle, as a function of radial camera distance $r_\text{cam}$ at fixed $N_\text{precision}=10^{-20}$ (left panel) and as a function of numerical precision $N_\text{precision}$ at fixed radial camera distance $r_\text{cam}=10^6\,r_g$ (right panel) for $a=0.99\,r_g$ in classical Kerr spacetime. Blue points show explicit numerical data points. For the radial distance, we also show the fitted function (red dashed), cf.~Eq.~\eqref{eq:error_deflectionAngle}.}
\end{figure}

\begin{figure}
\centering
\includegraphics[width=0.8\linewidth]{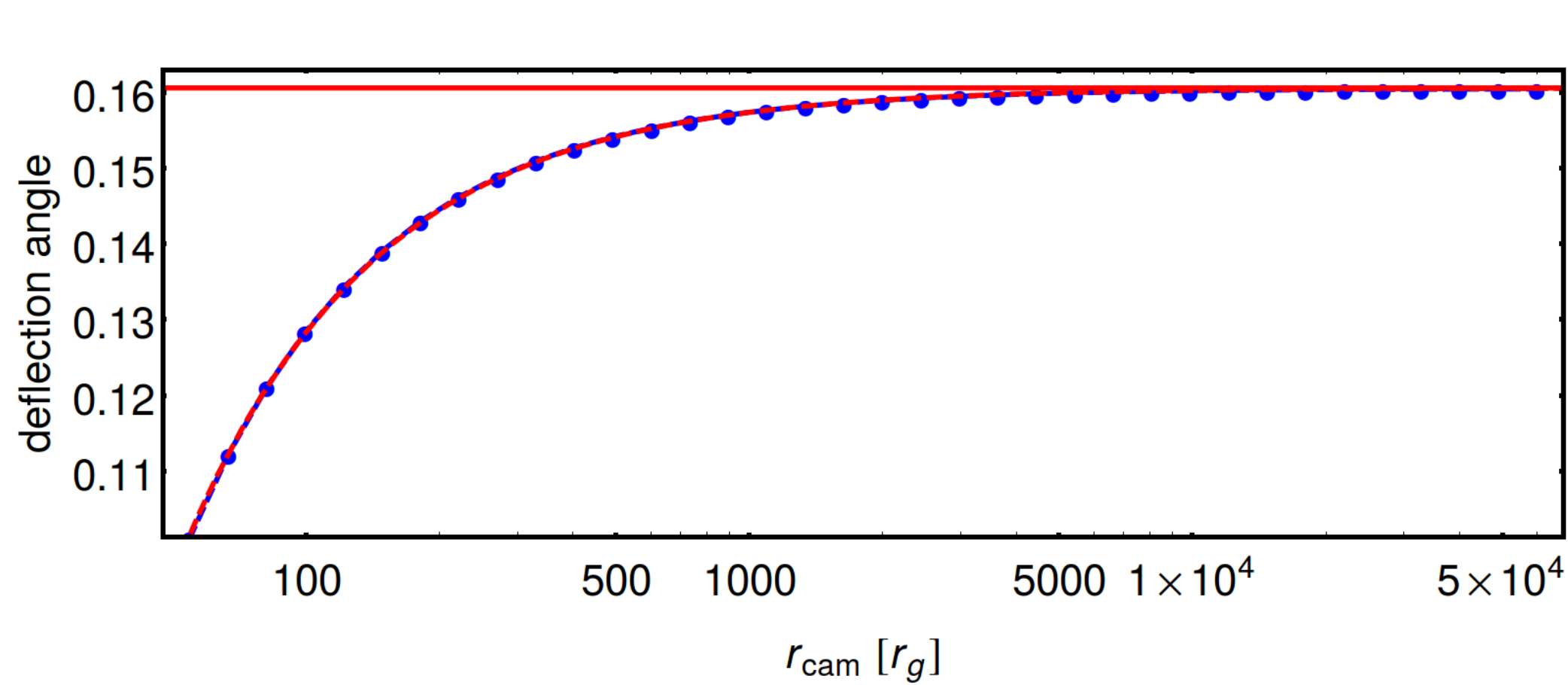}
\caption{\label{fig:errorKerrImproved}
Numerical data (blue connected points), fit (red-dashed) and fitted value $\vartheta_0$, cf.~Eq.~\eqref{eq:error_deflectionAngle} (red continuous) for the deflection angle in RG-improved Kerr spacetime with $a=0.9 r_g$ and $\tilde{\gamma} = 0.11$ at an impact parameter corresponding to the outermost image point $(x,y)=(10,0)$ in the equatorial plane. The plot shows the dependence on the radial distance $r_\text{cam}$ at fixed precision $N_\text{precision} = 10^{-20}$.
}
\end{figure}

Here, we demonstrate how to obtain the minimal required radial distance and numerical precision given a desired error-tolerance in the deflection angle in Kerr spacetime. As a point of reference, we choose the outermost image point $(x,y) = (10\, r_g,0)$ within the equatorial plane. Within the latter, the deflection angle is exactly calculable with arbitrary precision \cite{Iyer:2009wa}. Therefore, we use it as a benchmark test for the required numerical precision. figure~\ref{fig:errorKerr} shows how the numerical results (blue points) and the exact result (red continuous) develop with increasing radial distance and numerical precision. In both plots, we explicitly show how the error convergence stalls due to dominance of the respective other error. In practise, we avoid this by choosing both a large enough camera distance and numerical precision at the same time. Assuming that both errors do not significantly change for the RG-improved metric, we require a maximal error of the deflection angle $\Delta\vartheta = 0.001$ and therefore use $r_{\rm cam} = 10^4\, r_g$ and ${\rm Log}[N_\text{precision}] = -20$ throughout all computations.
Since the functional form, cf.~Eq.~\eqref{eq:error_deflectionAngle}, is known, we can test the convergence in radial distance in RG-improved Kerr spacetime explicitly, cf.~figure~\ref{fig:errorKerrImproved}.

\section{Features of the shadow for near-critical $\tilde{\gamma}$}
\label{app:kink_appendix}
\begin{figure}
\includegraphics[width=0.51\linewidth]{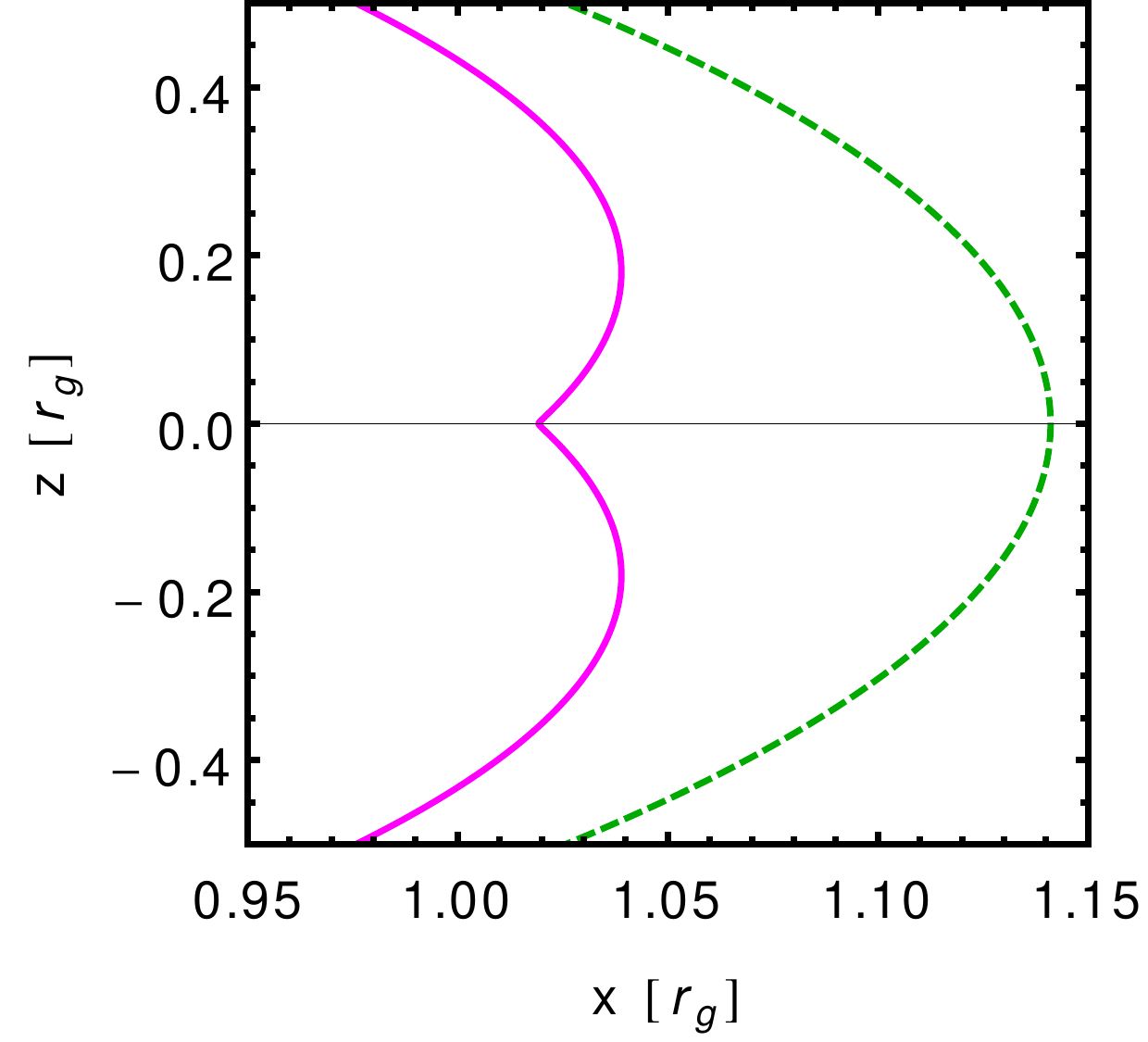}
\includegraphics[width=0.46\linewidth]{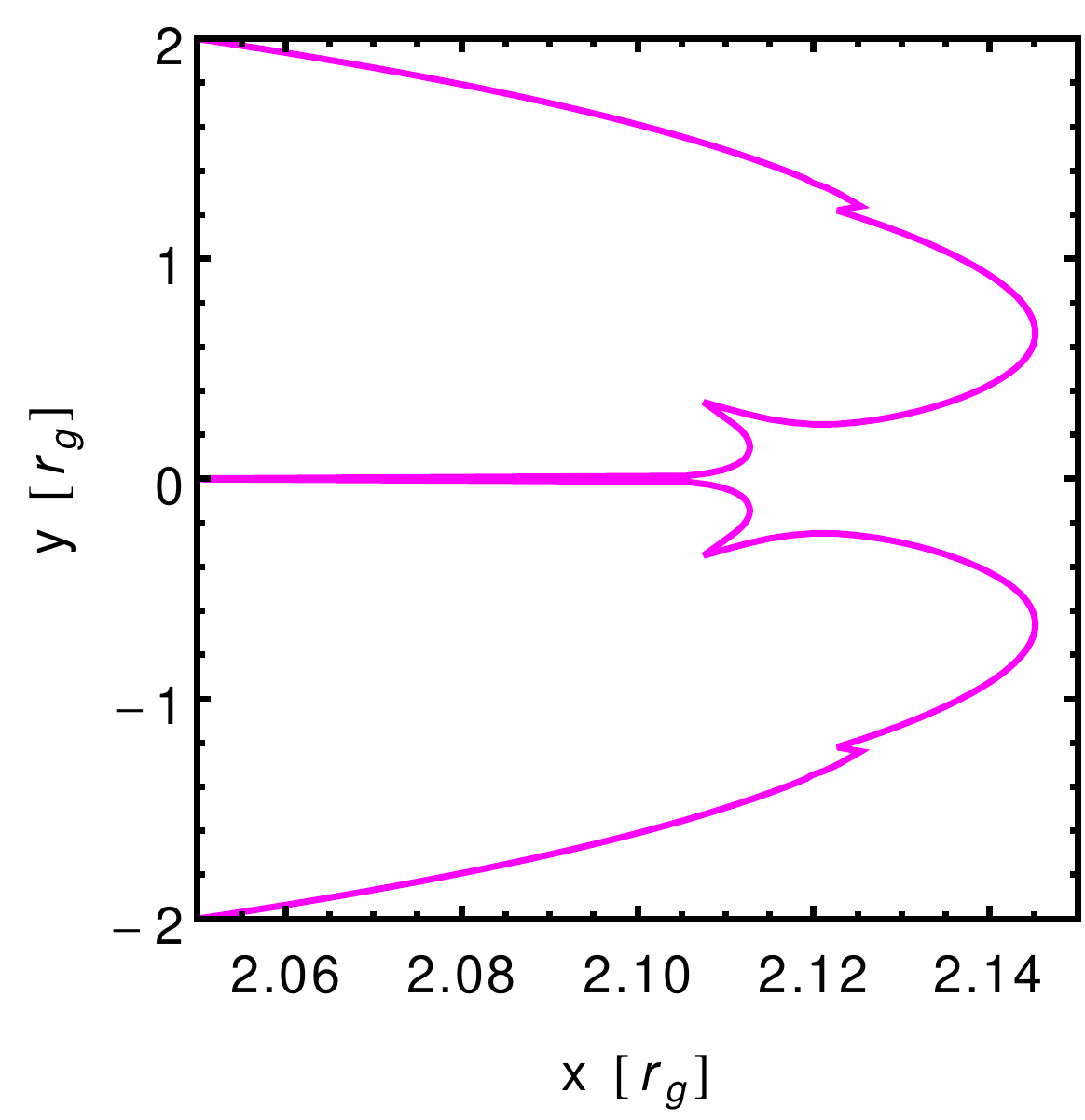}
\caption{\label{fig:extremeHorizonAndShadow}
Horizon for the classical and RG-improved Kerr spacetime (left panel) and the RG-improved shadow boundary (right panel) in the equatorial plane for a critical $\tilde\gamma=0.010242$ ($a=0.99 \, r_g$) just before the horizon disappears. 
}
\end{figure}

\begin{figure}
{\centering
\includegraphics[width=0.32\linewidth,trim={150pt 0 150pt 0},clip]{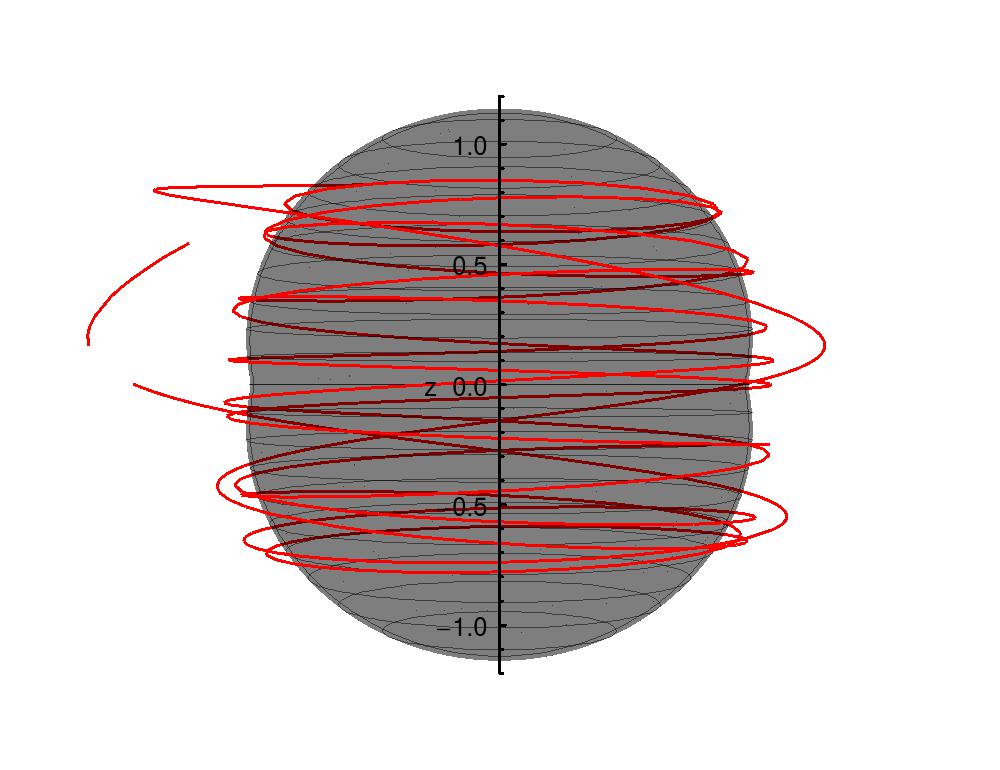}
\includegraphics[width=0.32\linewidth,trim={150pt 0 150pt 0},clip]{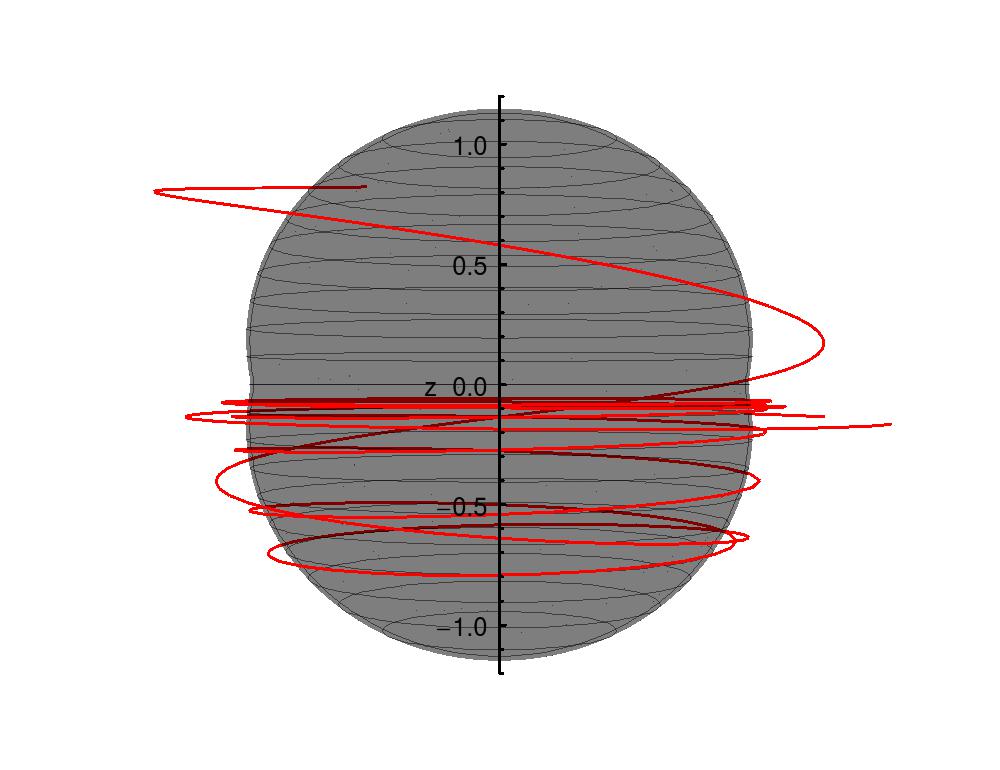}
\includegraphics[width=0.32\linewidth,trim={150pt 0 150pt 0},clip]{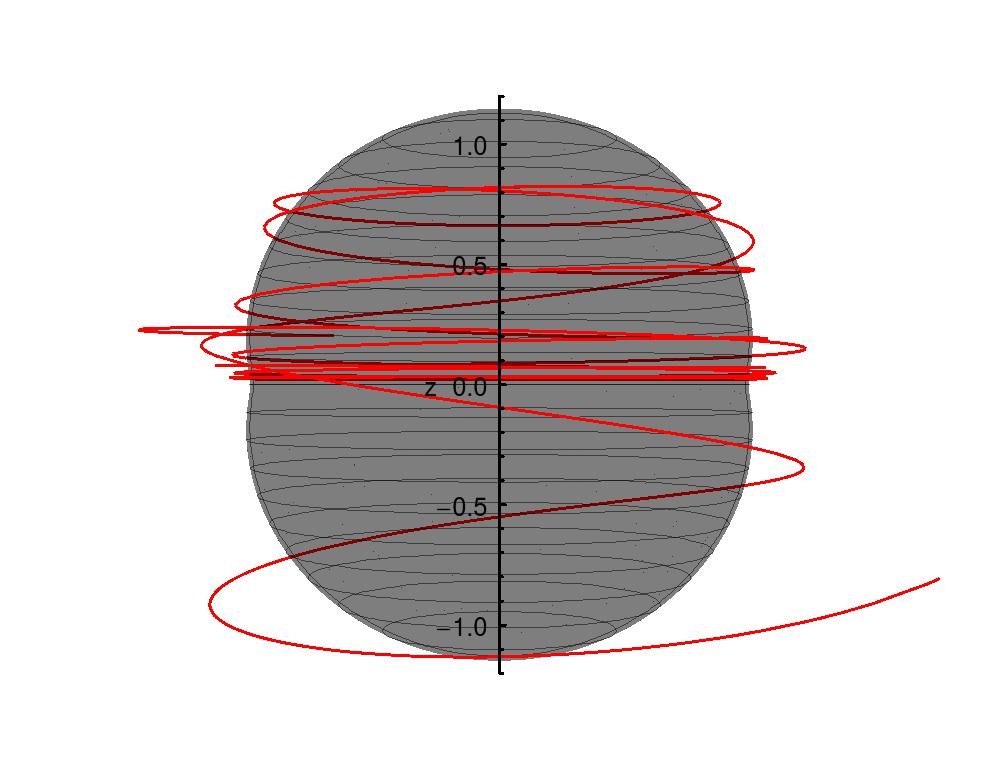}
}
\caption{\label{fig:extremeTrajectories}
Marginally stable light-like trajectories in the RG-improved black-hole spacetime (thick red lines) for an extreme $\tilde\gamma=0.010242$ ($a=0.99\, r_g$) 
just before the horizon (transparent surface) disappears. 
 The left panel shows an image angle $\psi>\psi_{\text{crit},\,1}$.
 The middle panel shows an image angle $\psi_{\text{crit},\,1}>\psi>\psi_{\text{crit},\,2}$.
 The right panel shows an image angle $\psi<\psi_{\text{crit},\,2}$.
}
\end{figure}

Close to critical $\tilde{\gamma}$, the horizon and the resulting shadow image develop additional distinct features.  Such features are non-generic  in the sense that they only appear for
 $\tilde{\gamma}$ close to  $\tilde{\gamma}_{\rm crit}$.
In the near-critical ($\tilde{\gamma}\approx \tilde{\gamma}_{\rm crit}$) regime, the dent in the  RG-improved horizon  at $\theta=\pi/2$ becomes very pronounced, cf.~left panel of figure~\ref{fig:extremeHorizonAndShadow}. This leads to a more pointy appearance of the dent-like feature in the shadow boundary at $\psi=0$.  Loosely speaking, the horizon takes on the appearance of two largely but not fully overlapping spheres (while remaining differentiable at $\theta=\pi/2$).
Further, it  results in  two sets of distinct, novel features at  two intermediate image angles, e.g.~for $a=0.99\, r_g$ and $\tilde{\gamma} \approx \tilde{\gamma}_\text{crit} = 0.010242$,  these occur at $\psi_{\rm crit,\,1}\approx \frac{16}{100}\pi$ and $\psi_{\rm crit,\,2}=\frac{4}{100}\pi$, cf.~right panel of figure~\ref{fig:extremeHorizonAndShadow}.
These secondary features  are a consequence of three different regimes for near-horizon null geodesics, cf.~figure~\ref{fig:extremeTrajectories}. For $\psi>\psi_{\text{crit},\,1}$ the light rays closest to the horizon probe the entire horizon.  At $|\psi| = \psi_{\text{crit},\,1}$, the null geodesics transition from wrapping around the entire horizon to wrapping around roughly half of the horizon. Loosely speaking, they probe just one of the two spheres that make up the horizon.
Accordingly, the shadow diameter grows significantly at $|\psi| \approx \psi_{\text{crit},\,1}$. In other words, a smooth, step-like feature appears in the shadow for $|\psi| = \psi_{\text{crit},\,1}$. As the dent in the shadow is rather prominent for $\tilde{\gamma}\approx \tilde{\gamma}_{\rm crit}$, it can "trap" trajectories that exist for $|\psi|< \psi_{\text{crit},\,2}$. As shown in figure~\ref{fig:extremeTrajectories}, these mainly wrap around the dented region of the horizon, and cover a significantly smaller interval in the affine parameter in exploring other parts of the horizon. Accordingly, these probe the smallest values of $r$ of all trajectories, and therefore arrive at values closer to the origin in the image plane.
 
For less extreme cases, i.e.~$\tilde{\gamma}< \tilde{\gamma}_{\rm crit}$, these features in the shadow-boundary become less pronounced. Nevertheless, traces of these features remain present in the shadow boundary. This can for instance be seen in figure~\ref{fig:radialdifference}  at $\psi\approx0.4$. 
\\
 We stress that we do not consider such features universal, in the sense that they can depend on the RG improvement that is used. The existence of the dent in the horizon and shadow, however, is robust.

\end{appendix}

\bibliography{References}
	
\end{document}